\documentclass[trackchanges,twocolumn]{aastex701}

\begin{document}

\title{Pre-perihelion Emergence of the CN Gas Coma in 3I/ATLAS Temporally and Spatially Resolved by the 7-Dimensional Telescope}

\author[orcid=0000-0002-6639-6533,gname='Gregory', sname='Paek']{Gregory S. H. Paek}  
\affiliation{Institute for Astronomy, University of Hawai`i, 2680 Woodlawn Drive, Honolulu, HI 96822, USA}
\email[show]{gregorypaek94@gmail.com} 

\author[orcid=0000-0002-8537-6714,gname=Myungshin, sname='Im']{Myungshin Im} 
\affiliation{Astronomy Program, Department of Physics and Astronomy, Seoul National University, Seoul 08826, Republic of Korea}
\affiliation{SNU Astronomy Research Center, Seoul National University, Seoul 08826, Republic of Korea}
\email[show]{myungshin.im@gmail.com}

\author[0009-0003-1280-0099]{Mankeun Jeong}
\affiliation{Astronomy Program, Department of Physics and Astronomy, Seoul National University, Seoul 08826, Republic of Korea}
\affiliation{SNU Astronomy Research Center, Seoul National University, Seoul 08826, Republic of Korea}
\email{jmk5040@gmail.com}

\author[0000-0003-4422-6426]{Hyeonho Choi}
\affiliation{Astronomy Program, Department of Physics and Astronomy, Seoul National University, Seoul 08826, Republic of Korea}
\affiliation{SNU Astronomy Research Center, Seoul National University, Seoul 08826, Republic of Korea}
\email{hhchoi1022@gmail.com}

\author[0000-0002-2618-1124]{Yoonsoo~P.~Bach}
\affiliation{Korea Astronomy and Space Science Institute (KASI), 776 Daedeok-daero, Yuseong-gu, Daejeon 34055, Republic of Korea}
\email{ysbach93@gmail.com}

\author[0000-0002-7332-2479]{Masateru Ishiguro}
\affiliation{Astronomy Program, Department of Physics and Astronomy, Seoul National University, Seoul 08826, Republic of Korea}
\affiliation{SNU Astronomy Research Center, Seoul National University, Seoul 08826, Republic of Korea}
\email{ishiguro@snu.ac.kr}

\author[0000-0002-8244-4603]{Bumhoo Lim}
\affiliation{Astronomy Program, Department of Physics and Astronomy, Seoul National University, Seoul 08826, Republic of Korea}
\affiliation{SNU Astronomy Research Center, Seoul National University, Seoul 08826, Republic of Korea}
\email{bumhoo7@snu.ac.kr}

\author[0000-0002-3118-8275]{Seo-Won Chang}
\affiliation{Astronomy Program, Department of Physics and Astronomy, Seoul National University, Seoul 08826, Republic of Korea}
\affiliation{SNU Astronomy Research Center, Seoul National University, Seoul 08826, Republic of Korea}
\email{seowon.chang@gmail.com}

\author[0000-0002-1418-3309]{Ji Hoon Kim}
\affiliation{Astronomy Program, Department of Physics and Astronomy, Seoul National University, Seoul 08826, Republic of Korea}
\affiliation{SNU Astronomy Research Center, Seoul National University, Seoul 08826, Republic of Korea}
\email{jhkim.astrosnu@gmail.com}

\author[0000-0002-3291-4056]{Jooyeon Geem}
\affiliation{Asteroid Engineering Laboratory, Space Systems, Lule\aa{} University of Technology, Box 848, SE-98128, Kiruna, Sweden}
\email{ksky0422@gmail.com}

\author[0000-0003-3953-9532]{Willem~B.~Hoogendam}
\altaffiliation{NSF Graduate Research Fellow}
\affiliation{Institute for Astronomy, University of Hawai`i, 2680 Woodlawn Drive, Honolulu, HI 96822, USA}
\email{willemh@hawaii.edu}

\begin{abstract}
We present time-series medium-band ($R\sim20$--40) observations of the third interstellar object 3I/ATLAS (C/2025 N1) obtained with the 7-Dimensional Telescope (7DT) which enables spatially resolved monitoring of its gas and dust activity from 2025 July to September. 
The m400-band image ($\lambda_c = 400$~nm, $\Delta\lambda \approx 25$~nm) reveals an emergence of a pronounced and spatially extended CN emission at a heliocentric distance of $r_h<3$~au. 
This onset is consistently identified across multiple diagnostics, including a break in the light-curve evolution, excess reflectance, inward expansion of annular excess beyond $10{,}000$--$20{,}000$~km, growth of the coma half-light radius from $\sim11{,}000$ to $\sim19{,}000$~km, and a rapid rise in the CN production rate $Q_{\rm CN}$ relative to $Af \rho$.
We further separate the CN-emitting and dust-scattered components via two-dimensional surface-brightness fitting into inner (dust) and outer (gas) components. 
We find that the outer component preserves a nearly constant profile shape, varying only in normalization, implying a relatively fast expansion of CN-bearing molecules.
Together, these results reveal how a transition occurs in the optical from dust-dominated scattering at large heliocentric distances to volatile-driven, gas-dominated activity as 3I/ATLAS enters the inner Solar System.
The timing and characteristics of the CN activation resemble the volatile enhancement observed in 2I/Borisov, suggesting that both known active interstellar objects exhibit comparable activation behavior at heliocentric distances of $\sim$2--3~au.
\end{abstract}




\section{Introduction}
Interstellar objects (ISOs) offer a rare window into the planetesimal reservoirs of extrasolar planetary systems. Since the discoveries of 1I/'Oumuamua (1I; \citealt{2017Natur.552..378M}) and 2I/Borisov (2I; \citealt{2020NatAs...4...53G}), the small ISO sample has already revealed striking diversity, from a coma-free object (1I) to a gas-rich comet (2I) with CN as the first and dominant optical volatile. 

The newly discovered 3I/ATLAS (C/2025 N1) on July 1, 2025, continues this diversity \citep{2025ApJ...989L..36S}. 
Rubin Observatory, ZTF, and NASA TESS precovery data revealed that 3I/ATLAS was already active at approximately 6.4~au as early as May 2025, with a dust-dominated coma and no detectable nucleus rotation, providing a crucial baseline for early activity prior to its discovery in July \citep{2025arXiv250713409C,2025ApJ...994L..51M,2025ApJ...991L...2F,2025ApJ...993L..31Y}.
The VLT/MUSE observations on 2025 July 3 (two days after discovery) showed a red, featureless continuum across 5000--9000\,\AA\ and no detected bands from CN, C$_2$, NH$_2$, or [O\,I], implying that the observed spectral reflectance at $r_h\simeq4.5$~au is predominantly governed by scattering from the dust coma \citep{2025MNRAS.544L..31O}.

However, by mid-August, independent optical campaigns reported the onset of CN emission.
A 10-night MDM program (Aug~8--17) captured the emergence and nightly strengthening of a weak CN feature near the 388.3~nm band head at $r_h \simeq 3.2$--$2.9$~au, deriving a CN production rate of $\log Q_{\mathrm{CN}} \sim 23$ (in molecules~s$^{-1}$) together with a stringent upper limit on the production-rate ratio of $\mathrm{C}_2$ to $\mathrm{CN}$, $\log[Q_{\mathrm{C}_2}/Q_{\mathrm{CN}}] < -1.05$, placing 3I among the class of carbon--chain--depleted comets \citep{2025ApJ...993L..23S}.
Complementary spectrophotometric observations with UH~2.2-m/SNIFS and integral-field spectroscopy with Keck-II/KCWI subsequently detected CN and Ni~\textsc{i} emission over multiple epochs between mid-August and early September, confirming a steady increase in CN production as the comet approached the Sun \citep{2025arXiv251209020H,2025arXiv251011779H}.
Near-simultaneous VLT X-shooter/UVES spectra independently detected CN and multiple Ni~\textsc{i} lines, likewise showing increasing production rates toward smaller heliocentric distances \citep{2025ApJ...995L..34R}.
Given the non-detection in early July \citep{2025A&A...700L..10A,2025arXiv251011779H}, these results indicate that CN outgassing switched on in mid-August.
This timing coincides with the detection of a CO$_2$-dominated coma: SPHEREx (Aug~12) detected strong CO$_2$ emission and water-ice absorption over a spatially extended coma \citep{2025arXiv251207318L}, and JWST/NIRSpec (Aug~18--19) revealed CO$_2$ with weak H$_2$O and CO \citep{2025ApJ...991L..43C}.

CN emission in comets originates primarily from photodissociation of parent molecules such as HCN released from the nucleus \citep{2004come.book..425F}. 
Observationally, CN is detected not only in the well-known violet system near 388~nm but also in a weaker red system at longer wavelengths near $\sim$787~nm; these bands arise from distinct electronic transitions of the CN radical and together provide complementary diagnostic information on the spatial distribution and excitation conditions of CN in cometary comae (e.g., \citealt{2004come.book..425F}).
However, spatially extended CN distributions observed in comets like 67P/Churyumov–Gerasimenko (67P) suggest an additional extended source, where CN parent molecules are preserved within dust grains and released via grain sublimation or fragmentation at distances $>10^4$~km from the nucleus \citep{2025P&SS..26806178M,2017MNRAS.469S.475R}. 
This gas-dust interaction--where volatiles are stored in and released from dust particles--plays a critical role in shaping the spatial structure and temporal evolution of coma activity. 
Understanding the spatial structure of CN emission therefore provides key constraints on the mechanisms of volatile release and the coupling between gas and dust in cometary comae.


Unlike 1I, which showed no detectable coma activity despite evidence for non-gravitational acceleration \citep{2018Natur.559..223M}, and 2I, for which CN emission was first detected only at relatively small heliocentric distances ($r_h \sim 2$~au) during its inbound evolution \citep{2020ApJ...889L..30L}, 3I/ATLAS exhibits a distinct activity pattern characterized by the onset of CN-dominated emission at larger heliocentric distances and persistent dust scattering throughout its approach to the Sun.
Beyond a simple census of dust and gas properties, it is essential to track how their temporal evolution as an ISO is progressively modified by the Solar-system environment, and to structurally resolve and deblend the coma and nucleus to isolate intrinsic activity from solar-system processing.

However, previous studies of ISO have provided limited spatially resolved, time-series coverage.
Therefore, time-series medium-band spectrophotometry with the 7-Dimensional Telescope (7DT; \citealt{2024SPIE13094E..0XK}) offers a unique dataset to understand their nature. 
It provides simultaneous coverage of $\sim$20–40 optical bands and a wide field of view, enabling low-resolution SEDs in what is effectively a ``spectroscopic mode,'' while spatially and temporally resolving the coma structure—an advantage over long-slit spectroscopy.
Leveraging annular photometry, we report not only the temporal evolution of SED but also CN line-dominated regions in projection.

We use 7DT observations of 3I/ATLAS from 2025 July 13 to September 18 ($r_h\simeq4.11$--$2.02$\,au) to 
(1) measure temporal brightening slopes across wavelength, 
(2) identify wavelength-dependent excesses diagnostic of composition, and 
(3) map their radial distributions to trace the CN gas and dust emission.
A pronounced m400-band enhancement, coinciding with the CN violet system near 388~nm (and potential C$_3$ near 405\,nm), allows us to trace the origin and spatial distribution of this blue-end excess and its implications for volatile release in extrasolar planetesimals.

\section{Observation}


We observed 3I/ATLAS with 7DT in spectroscopic mode over 14 epochs from 2025 July 13 to September 18 (UT). 
The system observes the target field with all 20 available medium-band filters and three broadband filters ($g$, $r$ and $i$). 
The medium-band set consists of 20 filters with 25 nm-bandwidths, spaced by 25 nm in central wavelength from 400 to 875 nm; each filter is named by its central wavelength with the prefix ``m'' (e.g., m400, spanning 387.5--412.5~nm). 
Per epoch, we obtained 3–18 frames per filter with 100~s on-source integrations. 
7DT was operated with the \texttt{TCSpy} program \citep{2024IAUGA..32P1281C,2024SPIE13101E..2VC}, which allows simultaneous control of the telescope units.


Because the number of available medium-band filters exceeds the sixteen telescope units, a full filter set for a given epoch could not be obtained strictly simultaneously.
Instead, subsets of filters were observed sequentially, resulting in a time offset of approximately $\sim$5 minutes between successive filter groups.
Given the characteristic rotation periods and variability timescales reported in previous studies ($\sim$16--17 hr; \citealt{2025A&A...700L...9D,2025A&A...702L...3S,2025ApJ...989L..36S,2025ApJ...991L...2F}), this offset is negligible for the purposes of our analysis.
We therefore treat the photometry obtained within each epoch as effectively simultaneous.

\begin{figure*}
    \centering
    \includegraphics[width=1\linewidth]{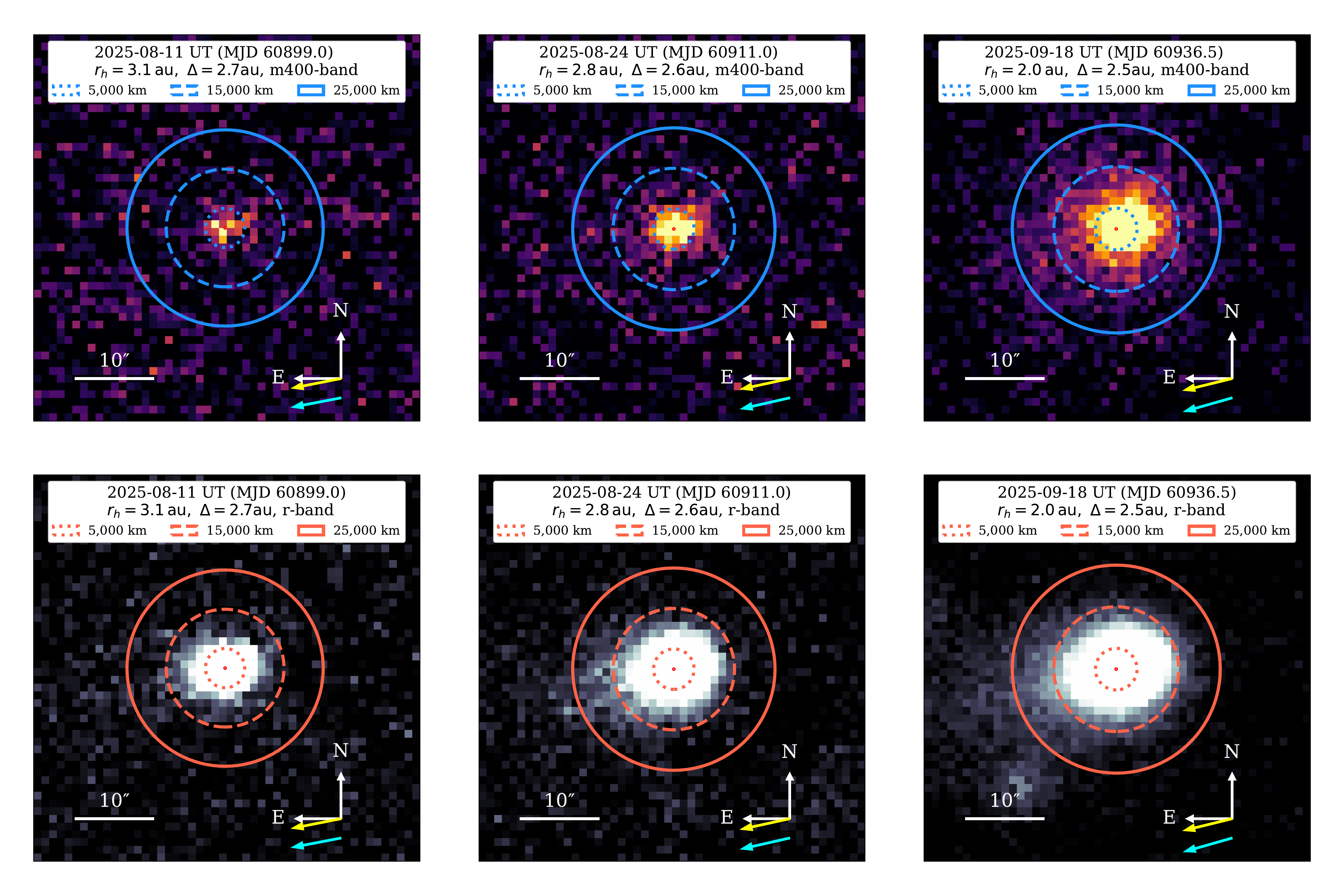}
    \caption{
$2\times2$-binned stacked images of 3I/ATLAS obtained with 7DT on 2025 August 11 and 24 and September 18 (UT), centered on the target. 
The top panel shows the m400 image, and the bottom panel shows the $r$-band image.
Concentric apertures corresponding to projected physical radius of $5{,}000$~km, $15{,}000$~km, and $25{,}000$~km are overplotted with dotted, dashed and solid lines for each; these were used to extract photometry at fixed physical scales given the changing geocentric distance.
White arrows indicate north and east (all panels share the same orientation), and yellow and cyan arrows mark the directions toward the Sun and of the object's motion, respectively.
}
    \label{fig:snapshot}
\end{figure*}

The observations were done with sidereal tracking, since the non-sidereal tracking mode of 7DT is still in the implementation stage. 
Nevertheless, 3I/ATLAS moves at about 2\arcsec\ over a 100~s frame while the apparent angular extent of the coma is of order several arcseconds (Figure~\ref{fig:snapshot}), enabling us to spatially resolve its extent. 
Table \ref{tab:data} shows the summary of the observation taken before the 3I/ATLAS's approach near the Sun prohibited observations in mid-September.

\section{Data}

\subsection{Preprocessing}
Each exposure was pre-processed and calibrated by the dedicated GPU pipeline \texttt{gpPy-GPU} \citep{paek2025_gppy_gpu}, which performs bias/dark/flat correction, astrometry, and zeropoint (ZP) calibration.
Because non-sidereal tracking was unavailable, we registered images using ephemerides from JPL HORIZONS \citep{1996DPS....28.2504G} at mid-exposure times and generated two products per filter/epoch: 
(1) a standard star-aligned median stack and 
(2) a target-centered median stack (Figure~\ref{fig:snapshot}). 
In crowded fields, we applied differential image analysis using a reference image from a pre-surveyed tile prior to stacking to mitigate background source confusion; both stacks share a consistent ZP solution.
The UT date, MJD, and heliocentric distance ($r_h$) corresponding to each epoch are indicated at the top of each panel.

\subsection{Photometric Calibration and Annular Measurements}
Photometric ZPs were derived from clean, unsaturated field stars by matching instrumental magnitudes to synthetic magnitudes computed from Gaia XP spectra \citep{2023A&A...674A...3M} convolved with 7DT-bandpasses. 
To minimize large-scale background systematics near the target, ZP calibrators were selected from within a circular aperture centered on the target, with a radius of $\sim$10$^\prime$.

Aperture photometry on the target-centered stacks was performed in fixed physical apertures with radii ($\rho$) sampled every $500$~km from $1{,}500$ to $25{,}000$~km. 
Annular (projected-radius) photometry was constructed by differencing adjacent physical apertures; these annuli form the basis of our reflectance and line-proxy maps (Section~\ref{subsec:lineproxy}). 
Solar-normalized reflectance spectra were computed per annulus by dividing calibrated fluxes by synthetic photometry of the Sun spectrum for each band (a high-resolution 
solar spectrum convolved with the 7DT throughput)\footnote{\url{https://www.stsci.edu/hst/instrumentation/reference-data-for-calibration-and-tools/astronomical-catalogs/solar-system-objects-spectra}}, and normalizing at m550.

\subsection{Geometry Normalization and Light-curve Evolution}

For inter-epoch comparison we normalized magnitudes to absolute magnitude $H(t)$ using the H--G phase function \citep{1989aste.conf..524B} adopting a canonical $G=0.15$ given our modest phase coverage per-band. 
We fitted the derived $H(t)$ with a weighted linear function to obtain per-band heliocentric and temporal slopes within each physical aperture and annulus. 

\subsection{Line-proxy Construction and Activity Diagnostics}\label{subsec:lineproxy}
To isolate the blue-end excess associated with the CN gas emission (388~nm), we extrapolated the local continuum near 400\,nm using neighboring medium-bands (m425, m450, m475, and m500 when available). 
A weighted linear fit in $f_\nu$ provided the continuum estimate at $\lambda=400$\,nm; subtracting this from the calibrated m400 flux yielded a line-proxy flux density.
We converted the photometry to $f_\lambda$ and computed band-integrated line–proxy flux $F_{\rm line}$ by integrating over a fixed 2.5\,nm window centered on the CN(0–0) feature, rather than the full effective bandwidth of m400. 
We adopt 2.5\,nm as a representative width because CN emission for 3I/ATLAS is typically 2–3\,nm wide \citep{2025ApJ...995L..34R,2025ApJ...993L..23S,2025arXiv251011779H,2025arXiv251209020H} noting that this window captures only a subset of the full CN (0–0) band structure.

The CN (0–0) violet band is dominated by a sharp P-branch band head at 388.3~nm, followed by a series of R-branch lines extending toward shorter wavelengths.
Because the m400 filter bandpass (387.5--412.5~nm) only partially overlaps the CN (0–0) band, primarily covering the P-branch band head while excluding most of the R-branch emission shortward of $\sim$387.5~nm, the measured line-proxy flux samples only a fraction of the total CN (0–0) band emission.
As a consequence, the CN fluorescence efficiency adopted here, computed for the full CN (0–0) band, introduces a systematic scaling uncertainty when applied to this partially sampled bandpass.
The resulting CN production rates should therefore be regarded as lower limits, and a systematic underestimation by a factor of a few relative to fully band-integrated spectroscopic measurements is expected.



We derived the CN production rate, $Q_\mathrm{CN}$, from the measured line flux ($F_{\rm line}$) using a Haser model \citep{1957BSRSL..43..740H}.
CN fluorescence efficiencies were computed via the Lowell Observatory fluorescence calculator API\footnote{\url{https://asteroid.lowell.edu/docs/comet}; see \citet{2022A&C....4100661M}}, which accounts for the Swings effect through the heliocentric radial velocity; for the adopted observing geometry, we used a CN fluorescence efficiency of $(L/N)_{\rm CN} = 2.6 \times 10^{-13}\ {\rm erg\ s^{-1}\ molecule^{-1}}$, scaled as $r_h^{-2}$. 
We adopted Haser model scale lengths of $1.3 \times 10^{4}$~km for the parent molecule HCN and $2.1 \times 10^{5}$~km for the daughter molecule CN at 1~au, as compiled by \citet{1995Icar..118..223A}. The scale lengths were scaled with heliocentric distance as $r_h^{2}$ and describe the radial fall-off of the Haser coma profile.
This scaling corresponds to the commonly adopted assumption of a constant gas expansion velocity (typically $v_{\rm exp} = 1$~km~s$^{-1}$) in Haser-model-based analyses, enabling direct comparison with previously published CN production rates. At large heliocentric distances, the actual gas expansion velocity may differ from 
the canonical 1~km~s$^{-1}$ assumption and can vary with $r_h$ and activity level.
As a dust proxy we computed $Af\rho$ \citep{1984AJ.....89..579A} from the $r$-band continuum flux. 
Finally, we use $Q/Af\rho$ as a gas-to-dust activity ratio for comparisons with cometary and ISO samples.

\section{Results}
The m400 and $r$ bands exhibit distinct morphologies (Figure~\ref{fig:snapshot}). 
In the $r$ band, the dust coma is concentrated within $\lesssim 15{,}000$~km from the nucleus and shows an asymmetric, sunward-directed tail; both the coma and tail grow in physical extent over time. 
By contrast, the m400 morphology remains largely quiescent prior to the middle of August, but by September 18 it develops into an extended, relatively symmetric
feature reaching out to $25{,}000$~km from the nucleus.

\begin{figure*}
    \centering
    \includegraphics[width=0.75\linewidth]{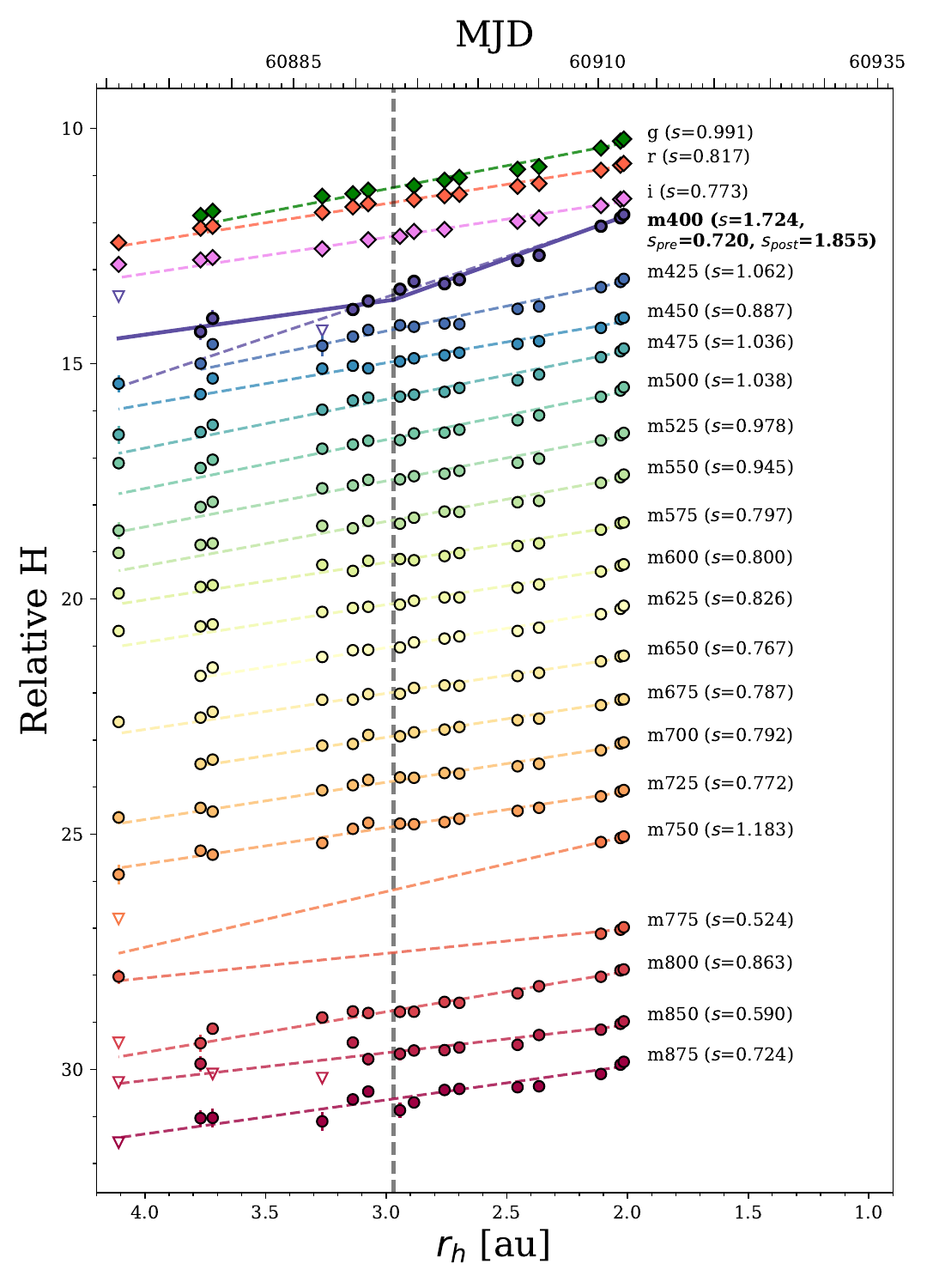}
    \caption{
Phase-- and distance--corrected absolute magnitudes $H$ of 3I/ATLAS as a function of heliocentric distance $r_h$ and MJD, measured within a projected $15{,}000$~km radius aperture.
The top axis shows the corresponding MJD.
For clarity, each filter sequence is vertically offset.
Filled circles (diamonds) denote detections ($S/N\ge5$) in medium bands (broad bands) with $1\sigma$ photometric uncertainties, while inverted triangles mark measurements with $S/N<5$ that are excluded from the fits.
Dashed lines show weighted linear fits of $H$ versus $r_h$ for each filter, with slopes $s$ (mag\,au$^{-1}$) printed next to the sequences.
For the m400 band, we additionally overplot a continuous broken--linear fit (solid line) with the break constrained to $2.5 \le x_b \le 3.5$~au and optimized from the data, yielding best--fit $x_b \simeq 2.97$~au, corresponding to the $\sim$MJD~60904.3 (vertical grey dashed line) and pre/post slopes ($s_{\rm pre}$ at larger $r_h$ and $s_{\rm post}$ at smaller $r_h$) as annotated.
}
    \label{fig:lc}
\end{figure*}


\begin{figure*}
    \centering
    \includegraphics[width=1.0\linewidth]{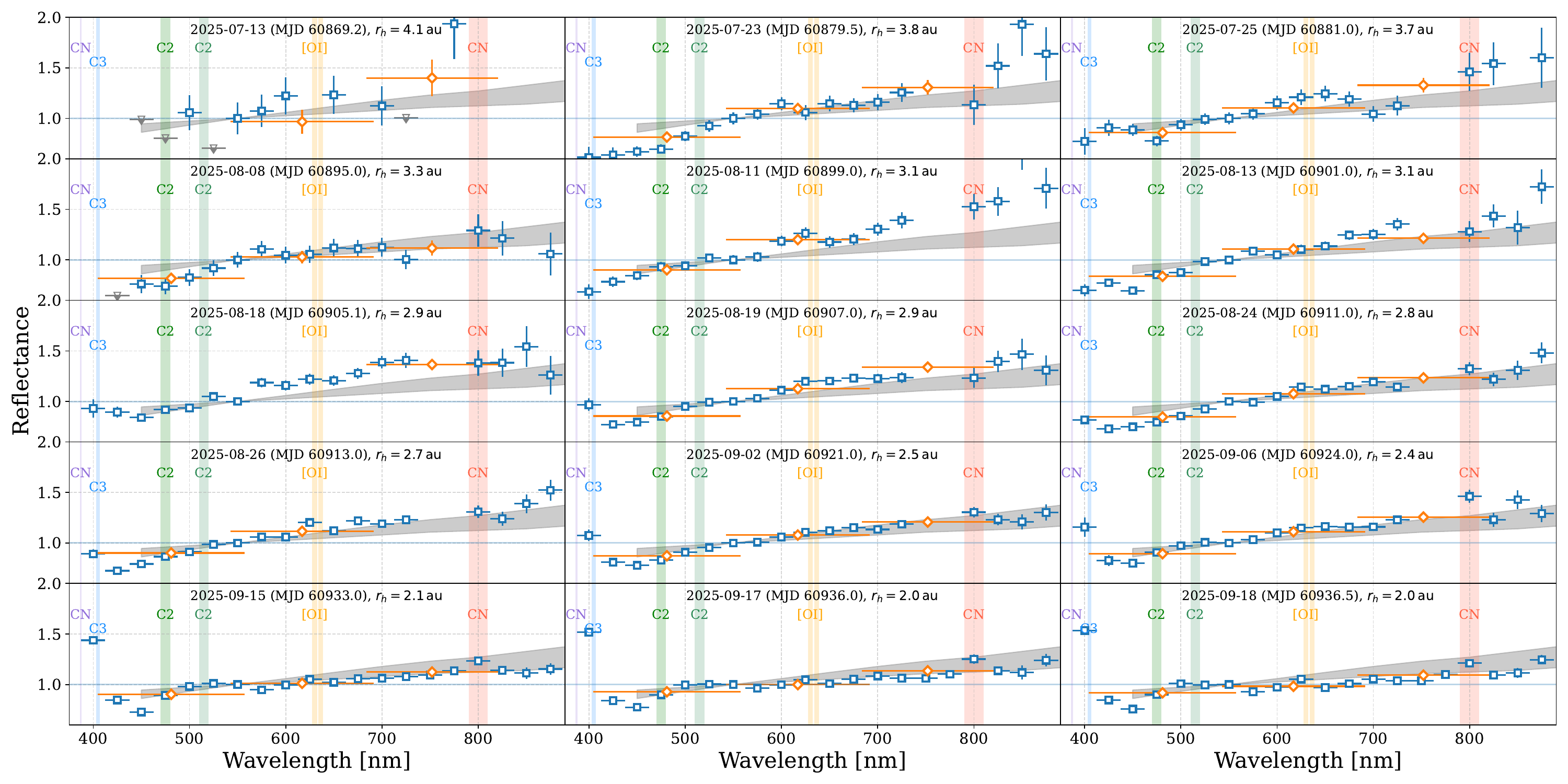}
    \caption{Time-series reflectance spectra of the interstellar object 3I/ATLAS obtained with 7DT medium-band (blue squares) and broadband (orange diamonds) photometry within circular apertures of radius $15{,}000$~km centered on the nucleus, obtained between 2025 July 13 and September 18. 
    Each panel corresponds to one observing night, with medium-band points normalized at the m550 filter. 
    Downward triangles indicate $5\sigma$ upper limits for non-detections. 
    The shaded gray region represents the reflectance template of D-type asteroids from \citet{2009Icar..202..160D} for reference. 
    Vertical colored-bands mark expected gas emission features: CN (violet and red), C$_2$ (green), C$_3$ (blue) and [O I] (yellow). 
}
    \label{fig:reflectance}
\end{figure*}

\subsection{Multi-band Light-Curve Brightening}\label{subsec:temporal_evolution}
Figure~\ref{fig:lc} shows the phase--distance corrected absolute magnitudes ($H$) of 3I/ATLAS as a function of heliocentric distance and time at different wavelengths, where $s$ denotes the slope of the light curve, expressed in units of mag~\,$au^{-1}$.
As the object approached perihelion, the brightness increased monotonically in all filters at a rate of about $0.89 \pm 0.24~\mathrm{mag~au^{-1}}$ ($-0.027 \pm 0.007~\mathrm{mag~day^{-1}}$).
This is in agreement with the post-transition activity slope ($-0.014$\,mag\,day$^{-1}$) after MJD~60870 (2025 July 14 UT) reported by \citet{2025ApJ...995L..15T}.



For the m400 band, a continuous broken--linear fit as a function of heliocentric distance ($r_h$) provides a significantly improved description of the data.
In this model, the breakpoint $x_b$ is defined as the heliocentric distance at which the slope of the light curve transitions between two linear regimes.
The best--fit breakpoint occurs at $r_h \simeq 2.97$~au, corresponding to MJD~60904.3 (2025 August 17 UT), and separates two distinct brightening regimes.
At larger heliocentric distances ($r_h > x_b$), the m400 brightness increases more gradually, with a slope of $s_{\rm pre} = 0.720~\mathrm{mag~au^{-1}}$, whereas at smaller distances ($r_h < x_b$) the brightening steepens markedly to $s_{\rm post} = 1.855~\mathrm{mag~au^{-1}}$.
The sharp change in slope across this distance strongly suggests the emergence of an additional blue--band component, most plausibly associated with enhanced gas emission, becoming dominant as the comet approaches the Sun.





The steeper brightening slope in m400 compared to other-bands can be interpreted as the onset of fluorescence emission from CN radicals. 
CN is produced through photodissociation of parent molecules (primarily HCN) by solar UV radiation \citep{2004come.book..425F}. 
In the cometary coma, CN fluorescence occurs when the radical absorbs solar photons and re-emits them at characteristic wavelengths near $388$~nm (the CN violet system). 
This process is distinct from dust scattering, which reflects sunlight across a broad continuum without spectral features. The localized enhancement in m400 therefore traces gas-phase emission rather than dust-dominated scattering.


%

The wavelength range of m400 encompasses potential contributions from C$_3$ emission near 405~nm as well as CN emission around 388~nm. 
However, previous spectroscopic observations around mid-August report no detectable C$_3$ emission, making it plausible that the observed m400 brightening is dominated by CN produced from HCN \citep{2025ApJ...995L..34R,2025ApJ...993L..23S,2025arXiv251209020H}, consistent with early upper limits on HCN followed by later detections at smaller heliocentric distances \citep{2025arXiv251202106H,2026MNRAS.tmp...52C}.

\subsection{Time-series Reflectance Spectra}

Figure~\ref{fig:reflectance} shows the temporal evolution of reflectance spectra, obtaining using a fixed aperture of 15{,}000~km. 
The rise of CN emission at 388~nm can be prominently identified in the spectra. 
We also see a weak excess in the m800 band after mid-August, which overlaps the wavelength region of the CN red system near 787~nm.
This is consistent with the CN emission onset reported by contemporaneous spectroscopy \citep{2025ApJ...993L..23S,2025arXiv251011779H}.
On the other hand, no other cometary emissions (e.g., C$_2$, C$_3$, or [O\,\textsc{i}]) can be seen.

\begin{figure}
    \centering
    \includegraphics[width=1.0\linewidth]{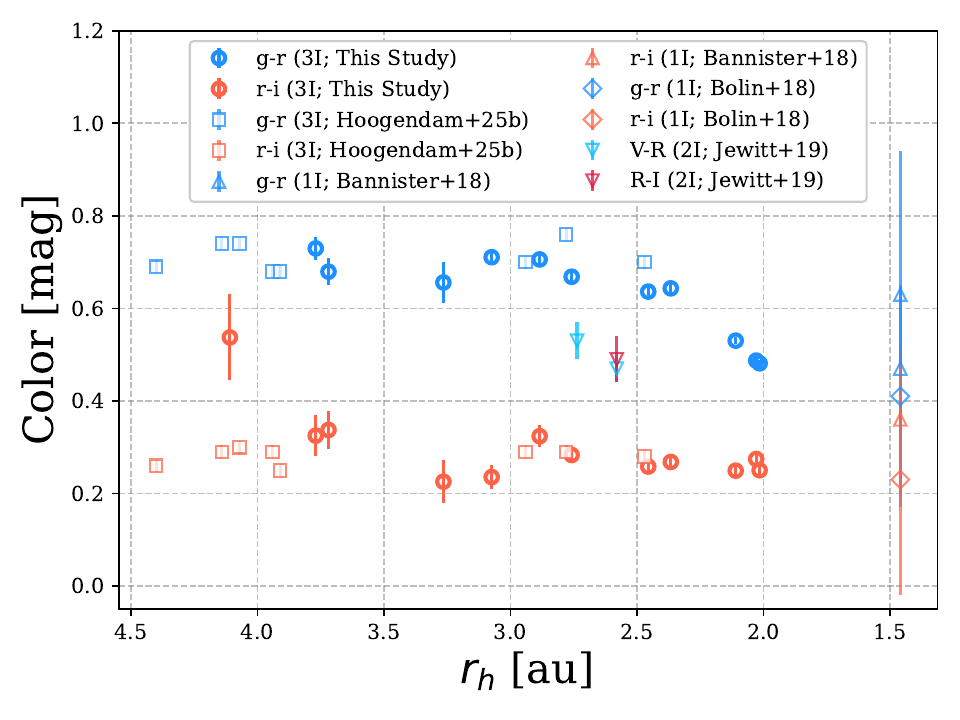}
    \caption{
Evolution of optical colors of 3I/ATLAS as a function of heliocentric distance.
Blue and red circles represent the $g-r$ and $r-i$ colors measured in this study, respectively, while open squares indicate early measurements from \citet{2025arXiv251209020H}.
For comparison, color measurements of the inactive ISO 1I \citep{2017ApJ...851L..38B,2018ApJ...852L...2B} and the active ISO 2I, \citep{2019ApJ...886L..29J} are also shown.
}
    \label{fig:color}
\end{figure}

The overall continuum shape is consistent with D-type mean reflectance in the Bus-Demeo taxonomy \citep{2009Icar..202..160D}, but slightly redder than that in July.
Figure~\ref{fig:color} shows the broadband color evolution of the continuum spectral slopes of 3I/ATLAS as summarized in Table~\ref{tab:data}.

\subsection{Gas and Dust Activity: $Q_{\rm CN}$ and $Af \rho$}\label{subsec:Q_Afp}

\begin{figure*}
    \centering
    \includegraphics[width=0.75\linewidth]{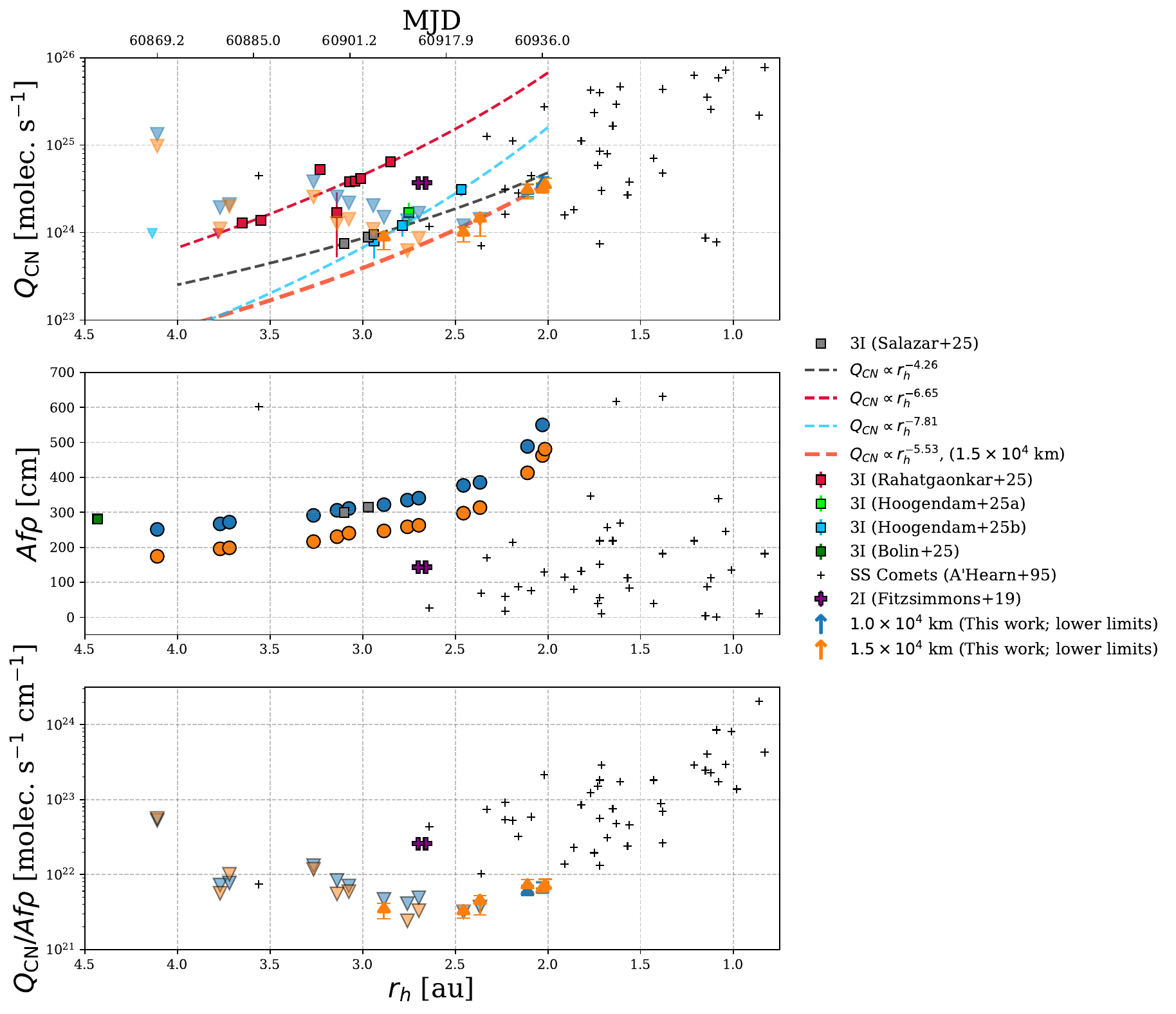}
    \caption{
    Time evolution of CN-related gas activity and dust production in 3I/ATLAS as a function of heliocentric distance $r_h$ (au); the upper abscissa shows the corresponding MJD based on our observing geometry.
    (Top) CN production rate $Q_{\rm CN}$ (molecules~s$^{-1}$);
    (Middle) $Af\rho$ (cm) as a proxy for dust production;
    (Bottom) the ratio $Q_{\rm CN}/Af\rho$ (molecules~s$^{-1}$,cm$^{-1}$).
    Our measurements are shown with $1\sigma$ uncertainties using fixed projected circular apertures of physical radius $1.0\times10^{4}$~km (blue upward arrows) and $1.5\times10^{4}$~km (orange upward arrows); the arrows indicate lower limits, while downward triangles mark $3\sigma$ upper limits.
    Dashed lines indicate representative power-law scalings of the form $Q_{\rm CN}\propto r_h^{\gamma}$ (see legend), including our fit for the $1.5\times10^{4}$~km aperture and literature-reported slopes for comparison.
    Black plus symbols show Solar System comets from \citet{1995Icar..118..223A}, evaluated at their reported $r_h$; these values were measured within a projected circular aperture of radius 10{,}000~km.
    Literature measurements for ISOs are distinguished by symbol shape.
    Gray squares show 3I/ATLAS from the MDM campaign \citep{2025ApJ...993L..23S}, where CN was extracted within a projected circular aperture of radius 15{,}000~km and $Af\rho$ was measured within 10{,}000~km.
    Red squares show 3I/ATLAS from VLT spectroscopy \citep{2025ApJ...995L..34R}, including X-shooter and UVES slit observations with effective projected slit lengths of order $(5$--$10)\times10^{3}$~km at the comet.
    Green and cyan squares indicate IFU-based extractions of 3I/ATLAS obtained with Keck-II/KCWI \citep{2025arXiv251011779H} and UH88/SNIFS \citep{2025arXiv251209020H}, respectively, using circular apertures of 2$^{\prime\prime}$ and 3$^{\prime\prime}$ radius, corresponding to projected physical radii of $\sim4\times10^{3}$ and $\sim6\times10^{3}$~km at $\Delta\sim2.5$--2.7~au.
    The purple cross denotes 2I from \citet{2019ApJ...885L...9F}, where CN was measured from long-slit spectroscopy and $Af\rho$ is commonly reported for a 10{,}000~km circular aperture.
    }
    \label{fig:Q_Afp}
\end{figure*}

Figure~\ref{fig:Q_Afp} presents the temporal evolution of both the CN production rate ($Q_{\rm CN}$) and the dust proxy $Af \rho$, measured using fixed projected circular apertures with radii of 10{,}000 and 15{,}000~km.
The 10{,}000~km aperture is commonly adopted in cometary studies, enabling direct comparison with previous measurements, while the 15{,}000~km aperture was chosen to encompass nearly the full extent of the m400 coma, as shown in Figure~\ref{fig:snapshot}.
Both quantities increase gradually as 3I/ATLAS approaches the Sun, followed by a noticeably steeper rise inside $r_h < 3$~au.

The coupled evolution of $Q_{\rm CN}$ and $Af \rho$ shows that gas-related activity strengthens more rapidly than the effective dust cross-section within $\sim$3~au.
A clear transition occurs near this heliocentric distance, where $Q_{\rm CN}$ exhibits a sharp acceleration accompanied by a break in slope, while $Af \rho$ continues to rise more modestly.
As a result, the $Q_{\rm CN}/Af\rho$ ratio increases toward smaller heliocentric distances, indicating that, relative to the dust continuum, CN emission becomes progressively more prominent at optical wavelengths as the comet enters the inner Solar System.
At the latest epochs (September 17 and 18), the $Q_{\rm CN}/Af\rho$ ratio does not show further rapid growth; however, the limited number of measurements prevents firm conclusions regarding a possible convergence toward equilibrium between gas production and effective dust cross-section.

Figure~\ref{fig:Q_Afp} also compares our measurements with representative literature data.
The MDM measurements \citep{2025ApJ...993L..23S}, which adopt a projected circular aperture of 15{,}000~km for CN extraction and 10{,}000~km for $Af\rho$,
as well as the Keck-II/KCWI \citep{2025arXiv251011779H} and UH88/SNIFS \citep{2025arXiv251209020H} IFU measurements, which use circular apertures corresponding to projected physical radii of a few $\times10^{3}$~km, are broadly consistent with our results within factors of a few.
In contrast, the VLT/X-Shooter and UVES measurements \citep{2025ApJ...995L..34R}, obtained using long-slit spectroscopy with effective projected slit lengths of order $(5$--$10)\times10^{3}$~km, yield CN production rates that are systematically higher by approximately an order of magnitude at comparable heliocentric distances.
In addition to aperture geometry, this systematic offset likely reflects the fact that our CN production rates are derived from a partial sampling of the CN (0–0) band using a m400 filter as described in Section~\ref{subsec:lineproxy}, whereas spectroscopic measurements integrate the full band emission.

Although the inferred power-law slopes of $Q_{\rm CN}$ as a function of $r_h$ span a relatively broad range among different studies, they all indicate a steep increase in CN-related activity toward smaller heliocentric distances.

The observed differences in absolute normalization and the modest variations in slope are most naturally attributed to differences in aperture size, aperture geometry, and observational methodology.
Our 7DT measurements employ fixed projected circular apertures with radii of 10{,}000 and 15{,}000~km, whereas spectroscopic observations obtained with long-slit or IFU instruments integrate CN emission over rectangular or otherwise non-circular effective apertures with different characteristic physical scales.

\subsection{Temporal Evolution of Reflectance Profile}\label{subsec:refletance}
\begin{figure*}
    \centering
    \includegraphics[width=1.0\linewidth]{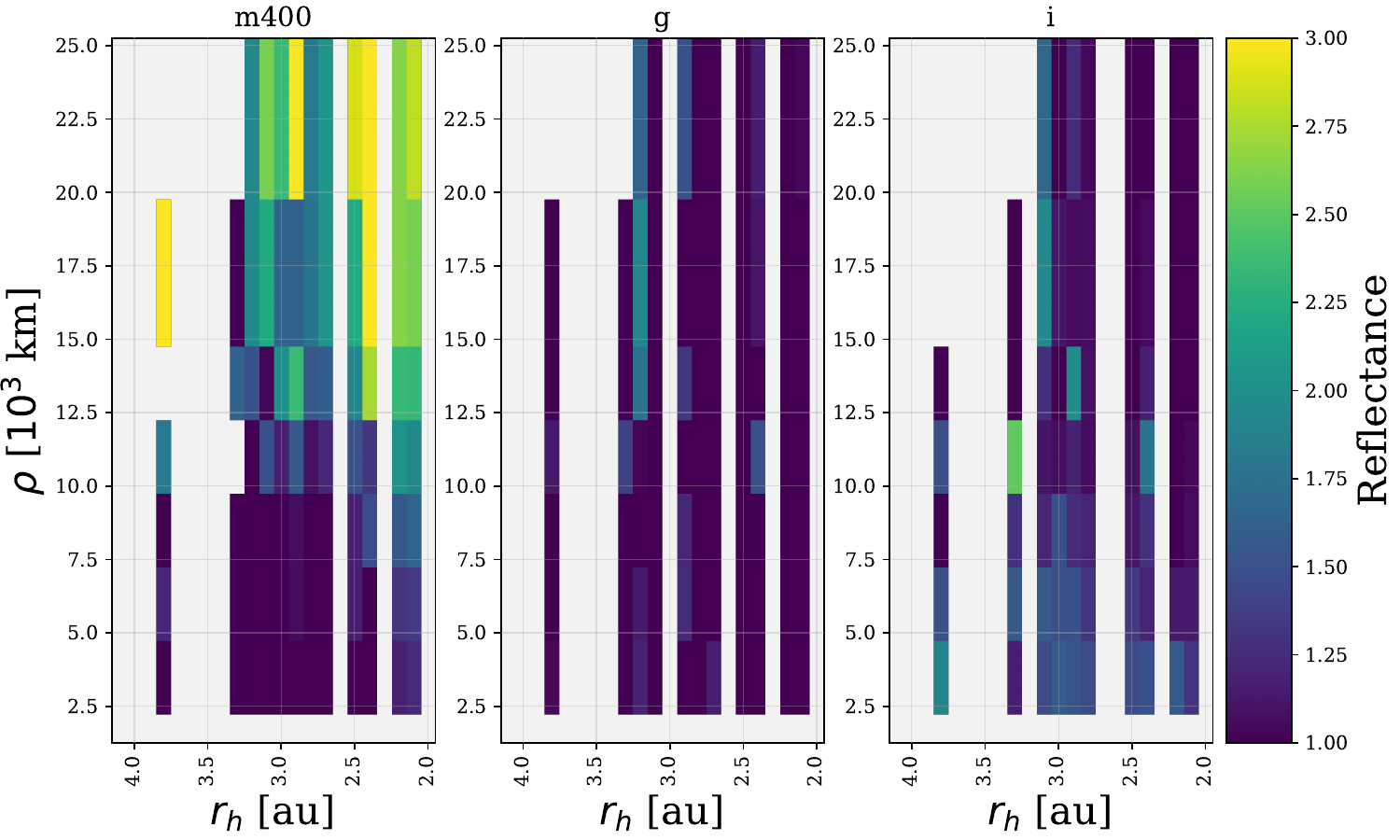}
    \caption{
    Temporal evolution of continuum-subtracted reflectance for 3I/ATLAS measured in the m400 band, which encompasses the CN emission feature.
    Colors represent annulus-averaged reflectance normalized to the m550 reflectance.
    The horizontal axis shows heliocentric distance ($r_{h}$), while the vertical axis indicates projected annuli ($\rho$) in physical units.
    The reflectance color scale is fixed to the range [1, 3] to facilitate inter-epoch comparison.
    Panels show, from left to right, the m400, $g$, and $i$ bands.
}

    \label{fig:m400_annulus}
\end{figure*}

We examined the radial distribution of reflectance in multiple filters, including m400, $r$, and $i$, using annular photometry (Figure~\ref{fig:m400_annulus}). 
In the m400 filter, an excess in reflectance appears in the coma at $r_h < 3.3$\,au, extending outward from $\sim 20{,}000$\,km. 
After the object reaches $r_h < 2.5$\,au, this excess becomes more pronounced in the outer coma and progressively extends inward toward the nucleus. 
In contrast, the $g$- and $i$-bands show nearly constant reflectances over all heliocentric distances and spatial scales, except perhaps a very marginal $i$-band reflectance excess within $\lesssim 15{,}000$\,km of the nucleus at $\sim 3$ au.
Other medium-band reflectance distribution shows no marked excess, similar to the $g$- and $i$-band cases.
These results reveal complementary spatial behaviors: with respect to dust-scattered light, the gas emission enhancement traced by m400 band occurs at larger coma radii.


\subsection{Two-Dimensional Surface Brightness Analysis Results}\label{subsec:profile}
\begin{figure*}
    \centering
    \includegraphics[width=0.75\linewidth]{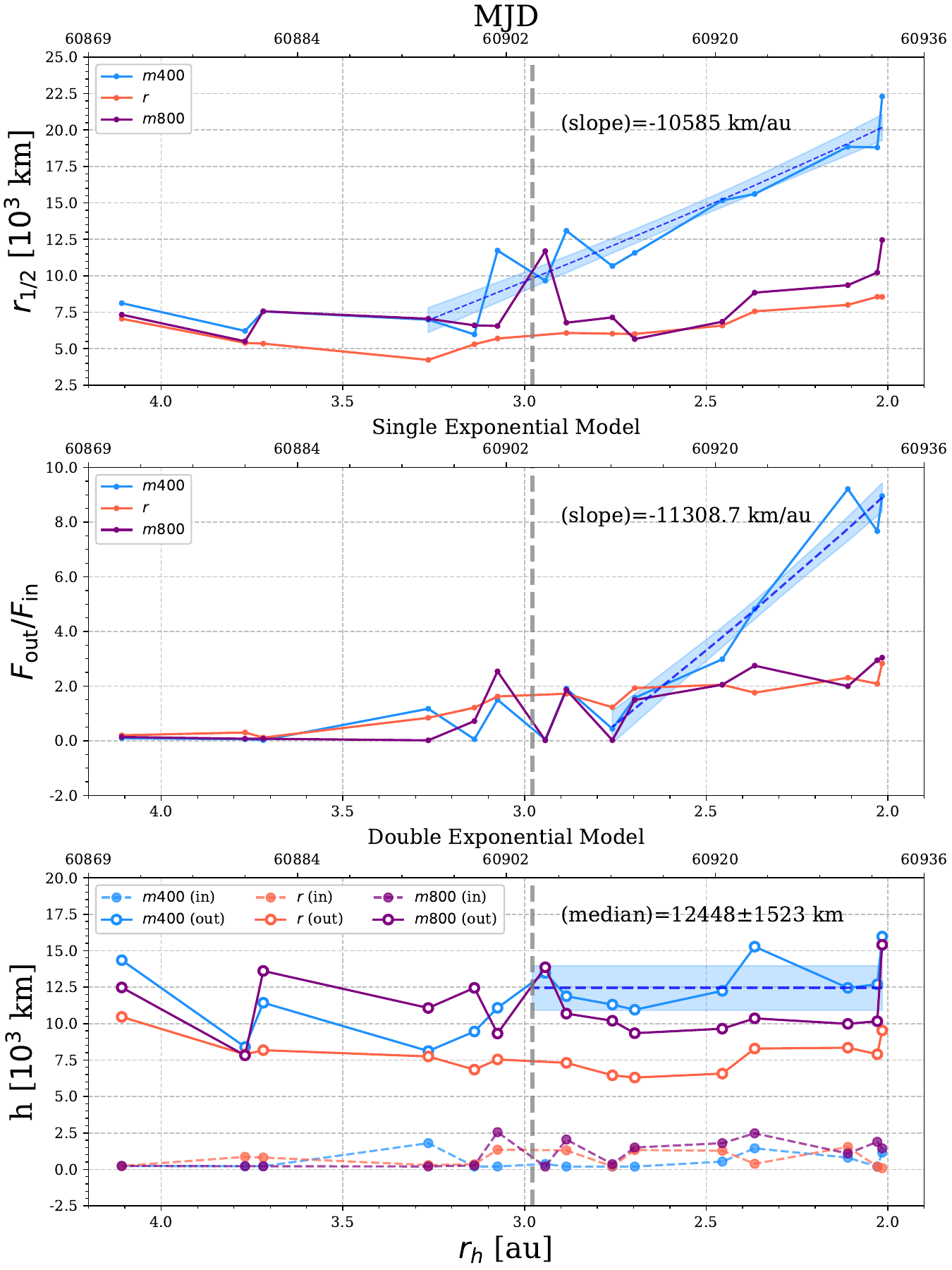}
    \caption{
The series of structural parameters and flux ratios as a function of time and heliocentric distance from \texttt{GALFIT} in physical units. 
\emph{Top}: half-light radius $r_{1/2}$ from the double-exponential model. 
\emph{Middle}: flux density ratio ($F_{\rm out}/F_{\rm in}$) between the two exponential components. 
\emph{Bottom}: characteristic scale lengths $h$ from the double-exponential model (inner: filled symbols; outer: open symbols). 
Colors denote filters: blue—m400 and red—$r$. 
The vertical gray dashed line marks MJD$=60904$. 
For m400, a linear regression over the post-break interval (MJD$>60904$) is shown in the top panel as a black dashed line with a 1$\sigma$ confidence band (blue shading); 
the best-fit slope is $d r_{1/2}/dr_h = -1.1\times10^{4}~\mathrm{km\,AU^{-1}}$. 
In the middle panel, the post-break median outer scale length for m400 is $\sim1.2\times10^4 \pm 1.5\times10^3$~km. 
The upper horizontal axis indicates the heliocentric distance $r_h$ (au).
}
    \label{fig:galfit}
\end{figure*}

We investigated the temporal evolution of the coma morphology in each filter by fitting the 2-dimensional (2D) surface brightness distribution using \texttt{GALFITM}\footnote{\url{https://www.nottingham.ac.uk/astronomy/megamorph/}} \citep{2013MNRAS.430..330H}, 
a multi-wavelength extension of \texttt{GALFIT}\footnote{\url{https://users.obs.carnegiescience.edu/peng/work/galfit/galfit.html}} \citep{2002AJ....124..266P,2010AJ....139.2097P} that enables simultaneous structural modeling across different bands.
As shown in Figure~\ref{fig:galfit}, we compared the m400 band---dominated by CN gas emission---with the $r$ band, which primarily traces dust continuum emission with minimal gas contamination.

The point spread function (PSF) was constructed using \texttt{PSFEx} \citep{2013ascl.soft01001B} from field stars with $S/N$~$>$~10, sampled across a $9\times9$ grid and normalized to unit flux at the comet's position. 
Background sources were masked using \texttt{photutils.segmentation} \citep{2021zndo...4624996B}, excluding pixels above 1.5$\sigma$ significance while retaining a central 100$\times$100 pixel region to avoid over-masking the coma. 


Each epoch was modeled using two concentric exponential disk components (inner and outer), together with a constant sky background. 
The centers of the two components were fixed to coincide, and the scale length (h) of the outer component was constrained to be larger than that of the inner component. 
This double-exponential representation is adopted as a phenomenological description of the radial surface brightness profile, serving as an empirical approximation to the Haser distribution \citep{1957BSRSL..43..740H} over the spatial scales probed by 7DT. 
In this framework, we interpret the inner component (with a shorter scale length) as tracing the dust-scattered continuum associated with the nucleus and inner coma, while the outer component represents the spatially extended coma. 
In particular, for the m400 band, the outer component is associated with CN-emitting gas that gives rise to the observed excess relative to the dust-dominated continuum.

The results of the fitting are shown in Figure~\ref{fig:galfit}, which shows the temporal evolution of half-light radii ($r_{1/2}$), scale lengths ($h$), and the flux density ratio of the outer to inner components ($F_{\rm out}/F_{\rm in}$)
We find that the $r_{1/2}$ value increases for m400 while it is nearly constant for the $r$ band, suggesting that m400 (the CN-dominated component) becomes more extended as the heliocentric distance decreases, whereas this behavior is not observed in the $r$ band (the dust-scattered component).
The middle panel shows the flux ratios ($F_{\rm in}/F_{\rm out}$) of the two components (outer to inner). 
For $r$ band, it is nearly $0$, meaning that the extended, outer component is almost negligible for the dust-scattered light. 
On the contrary, the outer component becomes significant starting from $~\sim3$~au, and this coincides with the photometric excess identified in Section~\ref{subsec:temporal_evolution}. 
Therefore, we consider that the outer component represents the CN excess for m400.
Regarding the scale lengths of the inner and outer components, the scale lengths ($h$) of the inner component are similar in the m400 and $r$ bands ($h \lesssim 1{,}000$~km), suggesting that the inner component traces the dust-scattered light distribution.
However, the scale length of the outer component in the $r$ band remains nearly constant, whereas that in the m400 band shows no marked evolution, indicating that the profile shape of the CN-emitting gas is conserved over the observed time span.
These results demonstrate that the onset of CN emission is driven by an increase in the flux of the outer component, rather than by an continuous increase in its spatial extent.

%

\section{Discussion}\label{sec:discussion}

\subsection{Emergence of Dust Rich, CN Emission Coma and Its Implication on Origin of 3I/ATLAS}\label{subsec:transition}

A broader picture for the origin of 3I/ATLAS emerges when comparing 3I/ATLAS with the ensemble of 45 Solar System comets from \citealt{1995Icar..118..223A}, as shown in Figure~\ref{fig:Q_Afp}. 
The CN production rates of 3I/ATLAS fall within the distribution spanned by typical Jupiter-family and long-period comets, indicating that its volatile output is not an outlier relative to the Solar System population.

The heliocentric dependence of $Q_{\rm CN}$ for 3I/ATLAS within a 15,000~km aperture follows a steep power-law slope of $\gamma~\sim-5.5$, consistent with the wide range of CN activity scalings ($Q_{CN} \propto r_h^{\gamma}$) observed among Solar System comets \citep{1995Icar..118..223A}.
In contrast, its dust production---as traced by $A f\rho$---is elevated by factors of several compared to most comets in the \citealt{1995Icar..118..223A} sample, placing 3I/ATLAS among the more dust-rich comae at comparable heliocentric distances.  
Consequently, the ratio $Q_{\rm CN}/A f\rho$ falls toward the lower envelope of the Solar System distribution: although not anomalously low, 3I/ATLAS exhibits a comparatively dusty coma for its level of CN production.

Contemporaneous spectroscopic observations show strong carbon-chain depletion, with upper limits on the C$_2$-to-CN ratio ($\log Q_{\mathrm{C}_2}/Q_{\mathrm{CN}} \lesssim -0.8$), placing 3I/ATLAS among the most carbon-chain depleted comets known \citep{2025ApJ...993L..23S,2025ApJ...995L..34R}. 
This combination of chemical signatures—high dust production, low relative CN output, and extreme carbon-chain depletion—provides important leverage on the formation environment and evolutionary history of 3I/ATLAS.

This behavior is consistent with the broader phenomenon of carbon-chain depletion seen in other comets, and discussed in detail by previous studies \citep[e.g.,][]{1995Icar..118..223A,2025ApJ...991L..43C}. 
In those works, carbon-chain depletion is generally attributed to a combination of formation in cold, UV-irradiated and/or metal-poor regions of the protoplanetary disk, where the build-up of carbon-chain precursors is inefficient, and subsequent long-term processing by UV photons and cosmic rays during residence in the outer system or interstellar space. 

For 3I/ATLAS, its likely low-metallicity birth environment and cosmic-ray contamination during a long interstellar travel time make both effects plausible. 
Recent JWST observations, which reveal extreme CO$_2$- and CO-rich volatile compositions and strong surface reddening, support the interpretation that the near-surface materials have been substantially altered by long-term cosmic-ray processing \citep{2025arXiv251026308M}. 
Additional constraints may come from the temporal evolution of volatile production and changes in the nucleus/coma structure around perihelion.

\subsection{Evolution of Spatial Distribution of CN and Dust Components}\label{subsec:structure}
The evolution of the spatial distributions of the CN and dust components shows a clear and physically interpretable segregation.  
Morphologically, the m400 band exhibits a nearly spherically symmetric coma extending beyond $25{,}000$~km (Figure~\ref{fig:galfit}), while the $r$ band displays a sunward-pointing dust tail. 


The structural properties of the m400-emitting gas morphology indicate that, although the spatial extent of the CN-emitting coma increases during the early phase of activation, its overall characteristic scale remains largely unchanged once CN emission becomes dominant, indicating that the large-scale structure of the gas coma is preserved over the later stages of our observations.
This behavior suggests that the observed increase in the m400 flux is not driven by continued structural broadening of the extended component, but rather by an increase in the relative brightness of the CN-dominated outer component with respect to the inner dust-scattered component.

The invariance of the radial profile shape of the extended outer component suggests that the expansion velocity of the CN parent molecules is relatively fast. 
If the outflow velocity were slow, the increasing volatile production toward smaller heliocentric distances would lead to a progressive buildup of material near the nucleus, resulting in a more centrally concentrated radial profile with a reduced scale length as 3I/ATLAS approaches the Sun. 
No such trend is observed. 
Instead, the persistence of a self-similar outer profile is consistent with a rapid expansion of the major parent species, including HCN and its daughter CN. 
This interpretation is independently supported by recent JCMT detections of HCN, which show a simple single-Gaussian line profile with an expansion velocity of $v_{\rm exp}\approx0.46$~km~s$^{-1}$ at $r_h\sim2$~au, indicative of a dynamically uniform and freely expanding coma \citep{2025arXiv251002817C}. 
Together, the optical CN morphology and millimeter HCN spectroscopy consistently point to relatively rapid gas expansion in the inner heliocentric regime ($\sim0.5$--1~km~s$^{-1}$).
While the persistence of a self-similar outer CN profile is consistent with rapid expansion, a distributed or diffuse source of CN, arising from parent molecules released from dust grains throughout the coma, could also provide an alternative explanation for the observed morphology.
In this context, the expansion velocity measured from millimeter spectroscopy ($v_{\rm exp}\approx0.46$~km~s$^{-1}$ at $r_h\sim2$~au) indicates that canonical constant-velocity assumptions commonly adopted in simple Haser models should be regarded as approximate.

\subsection{Flattening of the Continuum Slope and Resurfacing Process}
Beyond the spatial structure and kinematics of the gas coma, the optical data also reveal systematic evolution in the properties of the dust-scattered continuum.
As shown in Table~\ref{tab:data} and Figure~\ref{fig:color}, both the blue- and red-wavelength reflectance slopes, together with the $g-r$ color, evolve toward less red values as 3I/ATLAS approaches the Sun, while the $r-i$ color remains approximately constant.
This wavelength-dependent behavior suggests that the observed color evolution is not a uniform reddening or fading of the dust continuum, but instead reflects changes that preferentially affect shorter optical wavelengths.
Such wavelength-dependent color evolution may reflect multiple physical processes, including changes in surface properties as well as variations in the characteristic dust particle size distribution, as grain size can influence the wavelength dependence of optical scattering.

One possible interpretation is a resurfacing process, in which activity-driven dust emission exposes or blankets fresher, less-processed material, leading to a reduction of the red spectral slope, particularly at shorter wavelengths.
Similar color evolution has been reported for Solar System bodies with irradiation-processed surfaces, where the onset of activity can diminish ultra-red spectral characteristics \citep[e.g.,][]{2002AJ....123.1039J,2015AJ....150..201J}.
In this context, the contrast between the extremely red colors of the inactive 1I and the more comet-like colors observed for active ISOs such as 2I and 3I/ATLAS may reflect differences in surface processing and activity states.


\section{Conclusion}

We present a time-series analysis of the volatile and dust activity of 3I/ATLAS using medium-band photometry from the 7-Dimensional Telescope. 
Our data uniquely capture the emergence of the CN-dominated coma spatially and temporally. 
The m400-band excess, its large-scale symmetry, and the emergence of an extended m400 component out to $\sim$25{,}000~km at heliocentric distances of $r_h \lesssim 3$~au together reveal a rapid transition.
This volatile activation pattern is qualitatively closer to that of 2I than 1I.


A comparison with Solar System comets shows that, while the CN production rate of 3I/ATLAS lies within the normal cometary distribution, its dust production ($Af \rho$) is enhanced by factors of several, placing it among the dustiest comets known. 
The resulting low $Q_{\rm CN}/Af\rho$ ratio, combined with independently observed carbon-chain depletion, is consistent with previous findings for chemically depleted comets and offers a valuable opportunity to probe the origin as its volatile activity evolves through perihelion, underscoring the importance of continued 7DT monitoring.



The 2D analyses of the m400 surface brightness profile indicate the excess m400 emission arises mostly in the outer part of the comet, with the radial profile of the CN-dominated gas component becoming progressively more prominent relative to the dust component. 
In particular, the decomposition of the dust-component and CN gas component is achieved by the 2-dimensional surface brightness 
fitting with \texttt{GALFIT}. 
The \texttt{GALFIT} result shows that the inner component follows the evolution of dust components as found in other wavelengths, while the outer component becomes brighter as the comet nears the sun but preserves the surface brightness shape during this change. 
This interpretation is consistent with the relatively rapid expansion velocity ($\sim0.5$--1~km~s$^{-1}$) inferred for CN-bearing molecules and is independently supported by recent JCMT observations \citep{2025arXiv251002817C}.

The extended spatial distribution of CN emission, combined with its temporal acceleration inside $r_h\sim2.5$~au, provides strong evidence for a distributed source component in which CN parent molecules are released from dust grains rather than solely from the nucleus. 
Such gas--dust interaction mechanisms have been well established in several Solar System comets.
In particular, similar distributed CN sources have been identified in 67P \citep{2025P&SS..26806178M}, suggesting that this mechanism may represent a common process shared by ISOs and demonstrating that extended volatile sources are not confined to our Solar System.

While our observations capture the pre-perihelion activity of 3I/ATLAS, continued post-perihelion monitoring will be essential for fully characterizing the longevity and drivers of its volatile activity. 
Tracking the evolution of the spatial distributions and relative contributions of gas and dust components beyond perihelion will provide critical insight into the physical and compositional properties of this interstellar visitor. 
In this context, the time-series, wide-field spectral mapping achieved here with the 7DT demonstrates a unique, IFU-like capability that enables high-cadence monitoring of large-scale emission-line and continuum morphology. 
Applying this capability to future surveys will enable us to trace the evolutionary histories of both Solar System comets and rare interstellar visitors.


\begin{acknowledgments}

This work was supported by the National Research Foundation of Korea (NRF) grant, No. 2021M3F7A1084525, funded by the Korea government (MSIT).
G.S.H.P. acknowledges support from the Pan-STARRS project, which is a project of the Institute for Astronomy of the University of Hawai'i, and is supported by the NASA SSO Near Earth Observation Program under grants 80NSSC18K0971, NNX14AM74G, NNX12AR65G, NNX13AQ47G, NNX08AR22G, 80NSSC21K1572, and by the State of Hawai'i.
SWC acknowledges support from the Basic Science Research Program through the NRF funded by the Ministry of Education (RS-2023-00245013).
J.H.K. acknowledges the support from the Institute of Information \& Communications Technology Planning \& Evaluation (IITP) grant, No. RS-2021-II212068 funded by the Korean government (MSIT).
H.C. acknowledges the support from the National Research Foundation of Korea (NRF) grants, No. 2021M3F7A1084525 and RS-2025-00573214, funded by the Korean government (MSIT).
B.Lim acknowledges the support from the National Research Foundation of Korea (NRF) grants funded by the Ministry of Education (RS-2025-25436385)
W.B.H. acknowledges support from the National Science Foundation Graduate Research Fellowship Program under Grant No. 2236415. 
Data transfer from the host site was supported in part by the Korea Research Environment Open NETwork (KREONET) advanced research program funded by KISTI.
This job has made use of the Python package \texttt{GaiaXPy}, developed and maintained by members of the Gaia Data Processing and Analysis Consortium (DPAC), and in particular, Coordination Unit 5 (CU5), and the Data Processing Centre located at the Institute of Astronomy, Cambridge, UK (DPCI).






\end{acknowledgments}

\facilities{}

\software{Astropy \citep{2013A&A...558A..33A,2018AJ....156..123A,2022ApJ...935..167A},
          SciPy \citep{2020NatMe..17..261V},
          Source Extractor \citep{1996A&AS..117..393B}
          SWarp \citep{2010ascl.soft10068B},
          SCAMP \citep{2006ASPC..351..112B},
          PSFEx \citep{2013ascl.soft01001B},
          GALFITM \citep{2013MNRAS.430..330H},
          }




\begin{sidewaystable*}[t]
\caption{
Time-series m400 photometry of 3I/ATLAS.
The complete photometric dataset for all filters is provided as an electronic table.
\label{tab:data}
}
\centering
\begin{tabular}{ccccccccccccc}
\toprule
Date & MJD & Filter & $r$  & $\Delta$ & $\alpha$ & $H$   & $m$      & Refl. & Slope$_{\rm blue}$ & Slope$_{\rm red}$ & $g-r$ & $r-i$ \\
(UT) & (day) &      & (au) & (au)     & (deg)    & (mag) & (AB mag) &       & (\%/1000\AA)       & (\%/1000\AA)      & (mag) & (mag) \\
\midrule
2025-07-13 & 60869.2 & m400 & 4.11 & 3.17 & 6.2 & 12.07 & $18.13 \pm 0.33$ & $1.43 \pm 0.46$
& -- & -- & -- & $0.54 \pm 0.09$ \\

2025-07-23 & 60879.5 & m400 & 3.77 & 2.95 & 10.3 & 12.82 & $18.71 \pm 0.17$ & $0.61 \pm 0.10$
& $33.86 \pm 3.13$ & $14.03 \pm 2.27$ & $0.73 \pm 0.03$ & $0.33 \pm 0.04$ \\

2025-07-25 & 60881.0 & m400 & 3.72 & 2.92 & 11.0 & 12.54 & $18.40 \pm 0.18$ & $0.77 \pm 0.13$
& $14.75 \pm 3.44$ & $14.23 \pm 2.38$ & $0.68 \pm 0.03$ & $0.34 \pm 0.04$ \\

2025-08-08 & 60895.0 & m400 & 3.26 & 2.73 & 16.5 & 12.80 & $18.43 \pm 0.54$ & $0.43 \pm 0.21$
& $29.49 \pm 6.25$ & $6.85 \pm 2.36$ & $0.66 \pm 0.04$ & $0.23 \pm 0.05$ \\

2025-08-11 & 60899.0 & m400 & 3.14 & 2.69 & 18.0 & 12.35 & $17.91 \pm 0.11$ & $0.69 \pm 0.07$
& $14.34 \pm 2.52$ & $22.18 \pm 1.75$ & $0.72 \pm 0.02$ & -- \\

2025-08-13 & 60901.0 & m400 & 3.07 & 2.67 & 18.7 & 12.17 & $17.70 \pm 0.09$ & $0.70 \pm 0.06$
& $23.30 \pm 1.81$ & $16.50 \pm 1.38$ & $0.71 \pm 0.02$ & $0.24 \pm 0.03$ \\

2025-08-18 & 60905.1 & m400 & 2.94 & 2.64 & 20.0 & 11.92 & $17.37 \pm 0.10$ & $0.93 \pm 0.09$
& $11.53 \pm 2.36$ & $20.93 \pm 1.64$ & -- & -- \\

2025-08-19 & 60907.0 & m400 & 2.88 & 2.63 & 20.5 & 11.75 & $17.16 \pm 0.06$ & $0.97 \pm 0.06$
& $18.44 \pm 1.72$ & $15.41 \pm 1.26$ & $0.71 \pm 0.01$ & $0.32 \pm 0.02$ \\

2025-08-24 & 60911.0 & m400 & 2.76 & 2.61 & 21.5 & 11.80 & $17.13 \pm 0.05$ & $0.82 \pm 0.04$
& $24.78 \pm 1.24$ & $11.32 \pm 0.86$ & $0.67 \pm 0.01$ & $0.28 \pm 0.02$ \\

2025-08-26 & 60913.0 & m400 & 2.70 & 2.60 & 21.9 & 11.71 & $17.00 \pm 0.05$ & $0.89 \pm 0.04$
& $20.23 \pm 1.36$ & $13.72 \pm 0.88$ & $0.64 \pm 0.01$ & -- \\

2025-09-02 & 60921.0 & m400 & 2.46 & 2.57 & 23.0 & 11.31 & $16.40 \pm 0.05$ & $1.07 \pm 0.06$
& $19.75 \pm 1.47$ & $10.26 \pm 0.76$ & $0.64 \pm 0.01$ & $0.26 \pm 0.02$ \\

2025-09-06 & 60924.0 & m400 & 2.37 & 2.56 & 23.2 & 11.20 & $16.21 \pm 0.08$ & $1.16 \pm 0.09$
& $14.38 \pm 2.29$ & $12.92 \pm 0.95$ & $0.64 \pm 0.02$ & $0.27 \pm 0.02$ \\

2025-09-15 & 60933.0 & m400 & 2.11 & 2.54 & 22.6 & 10.58 & $15.31 \pm 0.02$ & $1.44 \pm 0.03$
& $17.25 \pm 0.85$ & $4.81 \pm 0.43$ & $0.53 \pm 0.01$ & $0.25 \pm 0.01$ \\

2025-09-17 & 60936.0 & m400 & 2.03 & 2.54 & 22.1 & 10.40 & $15.02 \pm 0.02$ & $1.52 \pm 0.04$
& $14.43 \pm 0.99$ & $4.55 \pm 0.49$ & $0.49 \pm 0.01$ & $0.27 \pm 0.01$ \\

2025-09-18 & 60936.5 & m400 & 2.02 & 2.53 & 22.0 & 10.33 & $14.93 \pm 0.02$ & $1.53 \pm 0.03$
& $14.77 \pm 0.81$ & $3.11 \pm 0.39$ & $0.48 \pm 0.01$ & $0.25 \pm 0.01$ \\
\bottomrule
\end{tabular}
\tablecomments{
$r$ and $\Delta$ denote heliocentric and geocentric distances, respectively.
$\alpha$ is the solar phase angle.
$H$ is the phase-corrected absolute magnitude.
Reflectance values are normalized at m550.
Slope values represent linear gradients measured in the blue and red wavelength ranges, with uncertainties quoted at the $1\sigma$ level.
Color indices ($g-r$ and $r-i$) are computed from matched aperture magnitudes; quoted uncertainties are $1\sigma$.
Missing color measurements are denoted by ``--''.
}
\end{sidewaystable*}


\bibliography{sample701}{}

@ARTICLE{2025arXiv251202106H,
       author = {{Hinkle}, Jason T. and {Yang}, Bin and {Meech}, Karen J. and {Hoffman}, Andrew and {Shappee}, Benjamin J. and {Hoogendam}, W.~B. and {Wray}, James J.},
        title = "{JCMT Constraints on the Early-Time HCN and CO Emission and HCN Temporal Evolution of 3I/ATLAS}",
      journal = {arXiv e-prints},
         year = 2025,
        month = dec,
          eid = {arXiv:2512.02106},
        pages = {arXiv:2512.02106},
          doi = {10.48550/arXiv.2512.02106},
archivePrefix = {arXiv},
       eprint = {2512.02106},
 primaryClass = {astro-ph.EP},
       adsurl = {https://ui.adsabs.harvard.edu/abs/2025arXiv251202106H}
}

@ARTICLE{2025ApJ...995L..34R,
       author = {{Rahatgaonkar}, Rohan and {Carvajal}, Juan Pablo and {Puzia}, Thomas H. and {Luco}, Baltasar and {Jehin}, Emmanuel and {Hutsem{\'e}kers}, Damien and {Opitom}, Cyrielle and {Manfroid}, Jean and {Aravind}, K. and {Marsset}, Micha{\"e}l and {Yang}, Bin and {Buchanan}, Laura and {Fraser}, Wesley C. and {Forbes}, John and {Bannister}, Michele and {Bodewits}, Dennis and {Bolin}, Bryce T. and {Belyakov}, Matthew and {Knight}, Matthew M. and {Snodgrass}, Colin and {Bufanda}, Erica and {Dorsey}, Rosemary and {Ferellec}, L{\'e}a and {La Forgia}, Fiorangela and {Lippi}, Manuela and {Murphy}, Brian and {Nayak}, Prasanta K. and {Vander Donckt}, Mathieu},
        title = "{Very Large Telescope Observations of Interstellar Comet 3I/ATLAS. II. From Quiescence to Glow: Dramatic Rise of Ni I Emission and Incipient CN Outgassing at Large Heliocentric Distances}",
      journal = {\apjl},
         year = 2025,
        month = dec,
       volume = {995},
       number = {1},
          eid = {L34},
        pages = {L34},
          doi = {10.3847/2041-8213/ae1cbc},
archivePrefix = {arXiv},
       eprint = {2508.18382},
 primaryClass = {astro-ph.SR},
       adsurl = {https://ui.adsabs.harvard.edu/abs/2025ApJ...995L..34R}
}

@ARTICLE{2025ApJ...995L..15T,
       author = {{Tonry}, John L. and {Denneau}, Jr., Larry and {Alarc{\'o}n}, Miguel R. and {Clocchiatti}, Alejandro and {Erasmus}, Nicolas and {Fitzsimmons}, Alan and {Licandro}, Javier and {Meech}, Karen J. and {Siverd}, Robert and {Weiland}, Henry},
        title = "{ATLAS Photometry of Interstellar Object 3I/ATLAS}",
      journal = {\apjl},
         year = 2025,
        month = dec,
       volume = {995},
       number = {1},
          eid = {L15},
        pages = {L15},
          doi = {10.3847/2041-8213/ae1f12},
archivePrefix = {arXiv},
       eprint = {2509.05562},
 primaryClass = {astro-ph.EP},
       adsurl = {https://ui.adsabs.harvard.edu/abs/2025ApJ...995L..15T}
}

@ARTICLE{2025ApJ...994L..51M,
       author = {{Martinez-Palomera}, Jorge and {Tuson}, Amy and {Hedges}, Christina and {Dotson}, Jessie and {Barclay}, Thomas and {Powell}, Brian},
        title = "{Prediscovery TESS Observations of Interstellar Object 3I/ATLAS}",
      journal = {\apjl},
         year = 2025,
        month = dec,
       volume = {994},
       number = {2},
          eid = {L51},
        pages = {L51},
          doi = {10.3847/2041-8213/ae1f91},
archivePrefix = {arXiv},
       eprint = {2508.02499},
 primaryClass = {astro-ph.EP},
       adsurl = {https://ui.adsabs.harvard.edu/abs/2025ApJ...994L..51M}
}

@ARTICLE{2025arXiv251209020H,
       author = {{Hoogendam}, W.~B. and {Kuesters}, D. and {Shappee}, B.~J. and {Aldering}, G. and {Wray}, J.~J. and {Yang}, B. and {Meech}, K.~J. and {Tucker}, M.~A. and {Huber}, M.~E. and {Auchettl}, K. and {Angus}, C.~R. and {Desai}, D.~D. and {Hinkle}, J.~T. and {Kiyokawa}, J. and {Paek}, G.~S.~H. and {Romagnoli}, S. and {Shi}, J. and {Syncatto}, A. and {Ashall}, C. and {Dixon}, M. and {Hart}, K. and {Hoffman}, A.~M. and {Jones}, D.~O. and {Medler}, K. and {Pfeffer}, C.},
        title = "{University of Hawaii 88-inch Telescope Observations of the Interstellar Comet 3I/ATLAS: Spectrophotometric Blue-Sensitive Spectral Time Series Spanning Two Months from Discovery}",
      journal = {arXiv e-prints},
         year = 2025,
        month = dec,
          eid = {arXiv:2512.09020},
        pages = {arXiv:2512.09020},
          doi = {10.48550/arXiv.2512.09020},
archivePrefix = {arXiv},
       eprint = {2512.09020},
 primaryClass = {astro-ph.EP},
       adsurl = {https://ui.adsabs.harvard.edu/abs/2025arXiv251209020H}
}

@ARTICLE{2025P&SS..26806178M,
       author = {{Murphy}, Brian P. and {Opitom}, Cyrielle and {Snodgrass}, Colin and {Deam}, Sophie E. and {Ferellec}, L{\'e}a and {Knight}, Matthew and {Okoth}, Vincent and {Yang}, Bin},
        title = "{Recent Chemo-morphological coma evolution of comet 67P/Churyumov─Gerasimenko}",
      journal = {\planss},
         year = 2025,
        month = nov,
       volume = {268},
          eid = {106178},
        pages = {106178},
          doi = {10.1016/j.pss.2025.106178},
archivePrefix = {arXiv},
       eprint = {2507.13979},
 primaryClass = {astro-ph.EP},
       adsurl = {https://ui.adsabs.harvard.edu/abs/2025P&SS..26806178M}
}

@ARTICLE{2025ApJ...993L..31Y,
       author = {{Ye}, Quanzhi and {Kelley}, Michael S.~P. and {Hsieh}, Henry H. and {Bellm}, Eric C. and {Chen}, Tracy X. and {Dekany}, Richard and {Drake}, Andrew and {Groom}, Steven L. and {Helou}, George and {Kulkarni}, Shrinivas R. and {Prince}, Thomas A. and {Riddle}, Reed},
        title = "{Prediscovery Activity of New Interstellar Object 3I/ATLAS: Rapid Brightening from 6 to 4 au}",
      journal = {\apjl},
         year = 2025,
        month = nov,
       volume = {993},
       number = {1},
          eid = {L31},
        pages = {L31},
          doi = {10.3847/2041-8213/ae147b},
archivePrefix = {arXiv},
       eprint = {2509.08792},
 primaryClass = {astro-ph.EP},
       adsurl = {https://ui.adsabs.harvard.edu/abs/2025ApJ...993L..31Y}
}

@ARTICLE{2025MNRAS.544L..31O,
       author = {{Opitom}, Cyrielle and {Snodgrass}, Colin and {Jehin}, Emmanuel and {Bannister}, Michele T. and {Bufanda}, Erica and {Deam}, Sophie E. and {Dorsey}, Rosemary C. and {Ferrais}, Marin and {Hmiddouch}, Said and {Knight}, Matthew M. and {Kokotanekova}, Rosita and {Leicester}, Brayden and {Marsset}, Micha{\"e}l and {Murphy}, Brian and {Okoth}, Vincent and {Ridden-Harper}, Ryan and {Vander Donckt}, Mathieu and {Ferellec}, L{\'e}a and {Hutsem{\'e}kers}, Damien and {Lippi}, Manuela and {Manfroid}, Jean and {Benkhaldoun}, Zouhair},
        title = "{Snapshot of a new interstellar comet: 3I/ATLAS has a red and featureless spectrum}",
      journal = {\mnras},
         year = 2025,
        month = nov,
       volume = {544},
       number = {1},
        pages = {L31-L36},
          doi = {10.1093/mnrasl/slaf095},
archivePrefix = {arXiv},
       eprint = {2507.05226},
 primaryClass = {astro-ph.EP},
       adsurl = {https://ui.adsabs.harvard.edu/abs/2025MNRAS.544L..31O}
}

@ARTICLE{2025ApJ...993L..23S,
       author = {{Salazar Manzano}, Luis E. and {Lin}, Hsing Wen and {Taylor}, Aster G. and {Seligman}, Darryl Z. and {Adams}, Fred C. and {Gerdes}, David W. and {Ruch}, Thomas and {Frincke}, Tessa T. and {Napier}, Kevin J.},
        title = "{Onset of CN Emission in 3I/ATLAS: Evidence for Strong Carbon-chain Depletion}",
      journal = {\apjl},
         year = 2025,
        month = nov,
       volume = {993},
       number = {1},
          eid = {L23},
        pages = {L23},
          doi = {10.3847/2041-8213/ae1232},
archivePrefix = {arXiv},
       eprint = {2509.01647},
 primaryClass = {astro-ph.EP},
       adsurl = {https://ui.adsabs.harvard.edu/abs/2025ApJ...993L..23S}
}

@ARTICLE{2025arXiv251011779H,
       author = {{Hoogendam}, W.~B. and {Shappee}, B.~J. and {Wray}, J.~J. and {Yang}, B. and {Meech}, K.~J. and {Ashall}, C. and {Desai}, D.~D. and {Hart}, K. and {Hinkle}, J.~T. and {Hoffman}, A. and {Hu}, E.~M. and {Jones}, D.~O. and {Medler}, K. and {Pfeffer}, C.},
        title = "{Spatial Profiles of 3I/ATLAS CN and Ni Outgassing from Keck/KCWI Integral Field Spectroscopy}",
      journal = {arXiv e-prints},
         year = 2025,
        month = oct,
          eid = {arXiv:2510.11779},
        pages = {arXiv:2510.11779},
          doi = {10.48550/arXiv.2510.11779},
archivePrefix = {arXiv},
       eprint = {2510.11779},
 primaryClass = {astro-ph.EP},
       adsurl = {https://ui.adsabs.harvard.edu/abs/2025arXiv251011779H}
}

@ARTICLE{2025A&A...702L...3S,
       author = {{Santana-Ros}, T. and {Ivanova}, O. and {Mykhailova}, S. and {Erasmus}, N. and {Kami{\'n}ski}, K. and {Oszkiewicz}, D. and {Kwiatkowski}, T. and {Hus{\'a}rik}, M. and {Ngwane}, T.~S. and {Penttil{\"a}}, A.},
        title = "{Temporal evolution of the third interstellar comet 3I/ATLAS: Spin, color, spectra, and dust activity}",
      journal = {\aap},
         year = 2025,
        month = oct,
       volume = {702},
          eid = {L3},
        pages = {L3},
          doi = {10.1051/0004-6361/202556717},
archivePrefix = {arXiv},
       eprint = {2508.00808},
 primaryClass = {astro-ph.EP},
       adsurl = {https://ui.adsabs.harvard.edu/abs/2025A&A...702L...3S}
}

@ARTICLE{2025arXiv251002817C,
       author = {{Coulson}, Iain M. and {Kuan}, Yi-Jehng and {Charnley}, Steven B. and {Cordiner}, Martin A. and {Chuang}, Yo-Ling and {Lee}, Yueh-Ning and {Lin}, Min-Kai and {Milam}, Stefanie N. and {Pimpanuwat}, Bannawit and {Roth}, Nathan X. and {{\.Z}{\'o}{\l}towski}, Micha{\l}},
        title = "{JCMT detection of HCN emission from 3I/ATLAS at 2.1 AU}",
      journal = {arXiv e-prints},
         year = 2025,
        month = oct,
          eid = {arXiv:2510.02817},
        pages = {arXiv:2510.02817},
          doi = {10.48550/arXiv.2510.02817},
archivePrefix = {arXiv},
       eprint = {2510.02817},
 primaryClass = {astro-ph.EP},
       adsurl = {https://ui.adsabs.harvard.edu/abs/2025arXiv251002817C}
}

@ARTICLE{2025ApJ...991L..43C,
       author = {{Cordiner}, Martin A. and {Roth}, Nathan X. and {Kelley}, Michael S.~P. and {Bodewits}, Dennis and {Charnley}, Steven B. and {Drozdovskaya}, Maria N. and {Farnocchia}, Davide and {Micheli}, Marco and {Milam}, Stefanie N. and {Opitom}, Cyrielle and {Schwamb}, Megan E. and {Thomas}, Cristina A. and {Bagnulo}, Stefano},
        title = "{JWST Detection of a Carbon-dioxide-dominated Gas Coma Surrounding Interstellar Object 3I/ATLAS}",
      journal = {\apjl},
         year = 2025,
        month = oct,
       volume = {991},
       number = {2},
          eid = {L43},
        pages = {L43},
          doi = {10.3847/2041-8213/ae0647},
archivePrefix = {arXiv},
       eprint = {2508.18209},
 primaryClass = {astro-ph.EP},
       adsurl = {https://ui.adsabs.harvard.edu/abs/2025ApJ...991L..43C}
}

@ARTICLE{2025arXiv251026308M,
       author = {{Maggiolo}, R. and {Dhooghe}, F. and {Gronoff}, G. and {de Keyser}, J. and {Cessateur}, G.},
        title = "{Interstellar Comet 3I/ATLAS: Evidence for Galactic Cosmic Ray Processing}",
      journal = {arXiv e-prints},
         year = 2025,
        month = oct,
          eid = {arXiv:2510.26308},
        pages = {arXiv:2510.26308},
          doi = {10.48550/arXiv.2510.26308},
archivePrefix = {arXiv},
       eprint = {2510.26308},
 primaryClass = {astro-ph.EP},
       adsurl = {https://ui.adsabs.harvard.edu/abs/2025arXiv251026308M}
}

@ARTICLE{2025ApJ...991L...2F,
       author = {{Feinstein}, Adina D. and {Noonan}, John W. and {Seligman}, Darryl Z.},
        title = "{Precovery Observations of 3I/ATLAS from TESS Suggest Possible Distant Activity}",
      journal = {\apjl},
         year = 2025,
        month = sep,
       volume = {991},
       number = {1},
          eid = {L2},
        pages = {L2},
          doi = {10.3847/2041-8213/adfd4d},
archivePrefix = {arXiv},
       eprint = {2507.21967},
 primaryClass = {astro-ph.EP},
       adsurl = {https://ui.adsabs.harvard.edu/abs/2025ApJ...991L...2F}
}

@ARTICLE{2025ApJ...989L..36S,
       author = {{Seligman}, Darryl Z. and {Micheli}, Marco and {Farnocchia}, Davide and {Denneau}, Larry and {Noonan}, John W. and {Hsieh}, Henry H. and {Santana-Ros}, Toni and {Tonry}, John and {Auchettl}, Katie and {Conversi}, Luca and {Devog{\`e}le}, Maxime and {Faggioli}, Laura and {Feinstein}, Adina D. and {Fenucci}, Marco and {Ferrais}, Marin and {Frincke}, Tessa and {Gillon}, Michael and {Hainaut}, Olivier R. and {Hart}, Kyle and {Hoffman}, Andrew and {Holt}, Carrie E. and {Hoogendam}, Willem B. and {Huber}, Mark E. and {Jehin}, Emmanuel and {Kareta}, Theodore and {Keane}, Jacqueline V. and {Kelley}, Michael S.~P. and {Lister}, Tim and {Mandt}, Kathleen and {Manfroid}, Jean and {Mar{\v{c}}eta}, Du{\v{s}}an and {Meech}, Karen J. and {Amine Miftah}, Mohamed and {Morgan}, Marvin and {Oca{\~n}a}, Francisco and {Pe{\~n}a-Asensio}, Eloy and {Shappee}, Benjamin J. and {Siverd}, Robert J. and {Taylor}, Aster G. and {Tucker}, Michael A. and {Wainscoat}, Richard and {Weryk}, Robert and {Wray}, James J. and {Yaginuma}, Atsuhiro and {Yang}, Bin and {Ye}, Quanzhi and {Zhang}, Qicheng},
        title = "{Discovery and Preliminary Characterization of a Third Interstellar Object: 3I/ATLAS}",
      journal = {\apjl},
         year = 2025,
        month = aug,
       volume = {989},
       number = {2},
          eid = {L36},
        pages = {L36},
          doi = {10.3847/2041-8213/adf49a},
archivePrefix = {arXiv},
       eprint = {2507.02757},
 primaryClass = {astro-ph.EP},
       adsurl = {https://ui.adsabs.harvard.edu/abs/2025ApJ...989L..36S}
}

@ARTICLE{2025A&A...700L..10A,
       author = {{Alvarez-Candal}, A. and {Rizos}, J.~L. and {Lara}, L.~M. and {Santos-Sanz}, P. and {Gutierrez}, P.~J. and {Ortiz}, J.~L. and {Morales}, N.},
        title = "{X-SHOOTER spectrum of comet 3I/ATLAS: Insights into a distant interstellar visitor}",
      journal = {\aap},
         year = 2025,
        month = aug,
       volume = {700},
          eid = {L10},
        pages = {L10},
          doi = {10.1051/0004-6361/202556338},
archivePrefix = {arXiv},
       eprint = {2507.07312},
 primaryClass = {astro-ph.EP},
       adsurl = {https://ui.adsabs.harvard.edu/abs/2025A&A...700L..10A}
}

@ARTICLE{2025A&A...700L...9D,
       author = {{de la Fuente Marcos}, R. and {Alarcon}, M.~R. and {Licandro}, J. and {Serra-Ricart}, M. and {de Le{\'o}n}, J. and {de la Fuente Marcos}, C. and {Lombardi}, G. and {Tejero}, A. and {Cabrera-Lavers}, A. and {Guerra Arencibia}, S. and {Ruiz Cejudo}, I.},
        title = "{Assessing interstellar comet 3I/ATLAS with the 10.4 m Gran Telescopio Canarias and the Two-meter Twin Telescope}",
      journal = {\aap},
         year = 2025,
        month = aug,
       volume = {700},
          eid = {L9},
        pages = {L9},
          doi = {10.1051/0004-6361/202556439},
archivePrefix = {arXiv},
       eprint = {2507.12922},
 primaryClass = {astro-ph.EP},
       adsurl = {https://ui.adsabs.harvard.edu/abs/2025A&A...700L...9D}
}

@ARTICLE{2025arXiv250713409C,
       author = {{Chandler}, Colin Orion and {Bernardinelli}, Pedro H. and {Juri{\'c}}, Mario and {Singh}, Devanshi and {Hsieh}, Henry H. and {Sullivan}, Ian and {Jones}, R. Lynne and {Kurlander}, Jacob A. and {Vavilov}, Dmitrii and {Eggl}, Siegfried and {Holman}, Matthew and {Spoto}, Federica and {Schwamb}, Megan E. and {Christensen}, Eric J. and {Beebe}, Wilson and {Roodman}, Aaron and {Lim}, Kian-Tat and {Jenness}, Tim and {Bosch}, James and {Smart}, Brianna and {Bellm}, Eric and {MacBride}, Sean and {Rawls}, Meredith L. and {Greenstreet}, Sarah and {Slater}, Colin and {Heinze}, Aren and {Ivezi{\'c}}, {\v{Z}}eljko and {Blum}, Bob and {Connolly}, Andrew and {Daues}, Gregory and {Makadia}, Rahil and {Gower}, Michelle and {Bryce Kalmbach}, J. and {Monet}, David and {Bannister}, Michele T. and {Dones}, Luke and {Dorsey}, Rosemary C. and {Fraser}, Wesley C. and {Forbes}, John C. and {Fuentes}, Cesar and {Holt}, Carrie E. and {Inno}, Laura and {Jones}, Geraint H. and {Knight}, Matthew M. and {Lintott}, Chris J. and {Lister}, Tim and {Lupton}, Robert and {Mendoza Magbanua}, Mark Jesus and {Malhotra}, Renu and {Mueller}, Beatrice E.~A. and {Murtagh}, Joseph and {Pandey}, Nitya and {Reach}, William T. and {Samarasinha}, Nalin H. and {Seligman}, Darryl Z. and {Snodgrass}, Colin and {Solontoi}, Michael and {Szab{\'o}}, Gyula M. and {White}, Ellie and {Womack}, Maria and {Young}, Leslie A. and {Allbery}, Russ and {Armellin}, Roberto and {Aubourg}, {\'E}ric and {Avdellidou}, Chrysa and {Azfar}, Farrukh and {Bauer}, James and {Bechtol}, Keith and {Belyakov}, Matthew and {Benecchi}, Susan D. and {Bertini}, Ivano and {Bolin}, Bryce T. and {Bose}, vMaitrayee and {Buchanan}, Laura E. and {Boucaud}, Alexandre and {Boufleur}, Rodrigo C. and {Boutigny}, Dominique and {Braga-Ribas}, Felipe and {Calabrese}, Daniel and {Camargo}, J.~I.~B. and {Caplar}, Neven and {Carry}, Benoit and {Carvajal}, Juan Pablo and {Choi}, Yumi and {Cowan}, Preeti and {Croft}, Steve and {{\'C}uk}, Matija and {Daruich}, Felipe and {Daubard}, Guillaume and {Davenport}, James R.~A. and {Daylan}, Tansu and {Delgado}, Jennifer and {Devillepoix}, Hadrien A.~R. and {Doherty}, Peter E. and {Donaldson}, Abbie and {Drass}, Holger and {Deppe}, Stephanie JH and {Dubois-Felsmann}, Gregory P. and {Economou}, Frossie and {Eduardo}, Marielle R. and {Farnocchia}, Davide and {Frissell}, Maxwell K. and {Fedorets}, Grigori and {Fernandes}, Maryann Benny and {Fulle}, Marco and {Gerdes}, David W. and {Gibbs}, Alex R. and {Gillan}, A. Fraser and {Guy}, Leanne P. and {Hammergren}, Mark and {Hanushevsky}, Andrew and {Hernandez}, Fabio and {Hestroffer}, Daniel and {Hopkins}, Matthew J. and {Granvik}, Mikael and {Ieva}, Simone and {Irving}, David H. and {Jannuzi}, Buell T. and {Jimenez}, David and {Ramos Gomes-J{\'u}nior}, Altair and {Juramy}, Claire and {Kahn}, Steven M. and {Kannawadi}, Arun and {Kang}, Yijung and {Kryszczy{\'n}ska}, Agnieszka and {Kotov}, Ivan and {Koumjian}, Alec and {Krughoff}, K. Simon and {Lage}, Craig and {Lange}, Travis J. and {Levine}, W. Garrett and {Li}, Zhuofu and {Licandro}, Javier and {Lin}, Hsing Wen and {Lust}, Nate B. and {Lyttle}, Ryan R. and {Mahabal}, Ashish A. and {Mahlke}, Max and {Plazas Malag{\'o}n}, Andr{\'e}s A. and {Salazar Manzano}, Luis E. and {Marc}, Moniez and {Margoti}, Giuliano and {Mar{\v{c}}eta}, Du{\v{s}}an and {Menanteau}, Felipe and {Meyers}, Joshua and {Mills}, Dave and {Morato}, Naomi and {More}, Surhud and {Morrison}, Christopher B. and {Moulane}, Youssef and {Mu{\~n}oz-Guti{\'e}rrez}, Marco A. and {Newcomer F.}, M. and {O'Connor}, Paul and {Oldag}, Drew and {Oldroyd}, William J and {O'Mullane}, William and {Opitom}, Cyrielle and {Oszkiewicz}, Dagmara and {Page}, Gary L. and {Patterson}, Jack and {Payne}, Matthew J. and {Peloton}, Julien and {Pereira}, Chrystian Luciano and {Peterson}, John R. and {Polin}, Daniel and {Pollek}, Hannah Mary Margaret and {Polen}, Rebekah and {Qiu}, Yongqiang and {Ragozzine}, Darin and {Rajagopal}, Jayadev and {van Reeven}, vWouter and {Rice}, Malena and {Ridgway}, Stephen T. and {Rivkin}, Andrew S. and {Robinson}, James E. and {Ro{\.z}ek}, Agata and {Salnikov}, Andrei and {S{\'a}nchez}, Bruno O. and {Sarid}, Gal and {Schambeau}, Charles A. and {Scolnic}, Daniel and {Schindler}, Rafe H. and {Seaman}, Robert and {Jacques}, {\v{S}}ebag and {Shaw}, Richard A. and {Shugart}, Alysha and {Sick}, Jonathan and {Siraj}, Amir and {Sitarz}, Michael C. and {Sobhani}, Shahram and {Soldahl}, Christine and {Stalder}, Brian and {Stetzler}, Steven and {Swinbank}, John D. and {Szigeti}, L{\'a}szl{\'o} and {Tauraso}, Michael and {Thornton}, Adam and {Tonietti}, Luca and {Trilling}, David E. and {Trujillo}, Chadwick A.},
        title = "{NSF-DOE Vera C. Rubin Observatory Observations of Interstellar Comet 3I/ATLAS (C/2025 N1)}",
      journal = {arXiv e-prints},
         year = 2025,
        month = jul,
          eid = {arXiv:2507.13409},
        pages = {arXiv:2507.13409},
          doi = {10.48550/arXiv.2507.13409},
archivePrefix = {arXiv},
       eprint = {2507.13409},
 primaryClass = {astro-ph.EP},
       adsurl = {https://ui.adsabs.harvard.edu/abs/2025arXiv250713409C}
}

@INPROCEEDINGS{2024SPIE13094E..0XK,
       author = {{Kim}, Ji Hoon and {Im}, Myungshin and {Lee}, Hyungmok and {Chang}, Seo-Won and {Choi}, Hyeonho and {Paek}, Gregory S.~H.},
        title = "{Introduction to the 7-Dimensional Telescope: commissioning procedures and data characteristics}",
    booktitle = {Ground-based and Airborne Telescopes X},
         year = 2024,
       editor = {{Marshall}, Heather K. and {Spyromilio}, Jason and {Usuda}, Tomonori},
       series = {Society of Photo-Optical Instrumentation Engineers (SPIE) Conference Series},
       volume = {13094},
        month = aug,
          eid = {130940X},
        pages = {130940X},
          doi = {10.1117/12.3019546},
       adsurl = {https://ui.adsabs.harvard.edu/abs/2024SPIE13094E..0XK}
}

@INPROCEEDINGS{2024IAUGA..32P1281C,
       author = {{Choi}, Hyeonho and {Im}, Myungshin and {Kim}, Ji Hoon},
        title = "{TCSpy: Multiple Telescope Control System for 7 Dimensional Telescope (7DT)}",
    booktitle = {32nd General Assembly International Union (IAUGA 2024)},
         year = 2024,
        month = aug,
          eid = {1281},
        pages = {1281},
       adsurl = {https://ui.adsabs.harvard.edu/abs/2024IAUGA..32P1281C}
}

@INPROCEEDINGS{2024SPIE13101E..2VC,
       author = {{Choi}, Hyeonho and {Im}, Myungshin and {Kim}, Ji Hoon},
        title = "{TCSpy: Multitelescope array control software for 7-Dimensional Telescope (7DT)}",
    booktitle = {Software and Cyberinfrastructure for Astronomy VIII},
         year = 2024,
       editor = {{Ibsen}, Jorge and {Chiozzi}, Gianluca},
       series = {Society of Photo-Optical Instrumentation Engineers (SPIE) Conference Series},
       volume = {13101},
        month = jul,
          eid = {131012V},
        pages = {131012V},
          doi = {10.1117/12.3018636},
       adsurl = {https://ui.adsabs.harvard.edu/abs/2024SPIE13101E..2VC}
}

@ARTICLE{2023A&A...674A...3M,
       author = {{Montegriffo}, P. and {De Angeli}, F. and {Andrae}, R. and {Riello}, M. and {Pancino}, E. and {Sanna}, N. and {Bellazzini}, M. and {Evans}, D.~W. and {Carrasco}, J.~M. and {Sordo}, R. and {Busso}, G. and {Cacciari}, C. and {Jordi}, C. and {van Leeuwen}, F. and {Vallenari}, A. and {Altavilla}, G. and {Barstow}, M.~A. and {Brown}, A.~G.~A. and {Burgess}, P.~W. and {Castellani}, M. and {Cowell}, S. and {Davidson}, M. and {De Luise}, F. and {Delchambre}, L. and {Diener}, C. and {Fabricius}, C. and {Fr{\'e}mat}, Y. and {Fouesneau}, M. and {Gilmore}, G. and {Giuffrida}, G. and {Hambly}, N.~C. and {Harrison}, D.~L. and {Hidalgo}, S. and {Hodgkin}, S.~T. and {Holland}, G. and {Marinoni}, S. and {Osborne}, P.~J. and {Pagani}, C. and {Palaversa}, L. and {Piersimoni}, A.~M. and {Pulone}, L. and {Ragaini}, S. and {Rainer}, M. and {Richards}, P.~J. and {Rowell}, N. and {Ruz-Mieres}, D. and {Sarro}, L.~M. and {Walton}, N.~A. and {Yoldas}, A.},
        title = "{Gaia Data Release 3. External calibration of BP/RP low-resolution spectroscopic data}",
      journal = {\aap},
         year = 2023,
        month = jun,
       volume = {674},
          eid = {A3},
        pages = {A3},
          doi = {10.1051/0004-6361/202243880},
archivePrefix = {arXiv},
       eprint = {2206.06205},
 primaryClass = {astro-ph.IM},
       adsurl = {https://ui.adsabs.harvard.edu/abs/2023A&A...674A...3M}
}

@ARTICLE{2022A&C....4100661M,
       author = {{Moskovitz}, N.~A. and {Wasserman}, L. and {Burt}, B. and {Schottland}, R. and {Bowell}, E. and {Bailen}, M. and {Granvik}, M.},
        title = "{The astorb database at Lowell Observatory}",
      journal = {Astronomy and Computing},
         year = 2022,
        month = oct,
       volume = {41},
          eid = {100661},
        pages = {100661},
          doi = {10.1016/j.ascom.2022.100661},
archivePrefix = {arXiv},
       eprint = {2210.10217},
 primaryClass = {astro-ph.EP},
       adsurl = {https://ui.adsabs.harvard.edu/abs/2022A&C....4100661M}
}

@ARTICLE{2022ApJ...935..167A,
       author = {{Astropy Collaboration} and {Price-Whelan}, Adrian M. and {Lim}, Pey Lian and {Earl}, Nicholas and {Starkman}, Nathaniel and {Bradley}, Larry and {Shupe}, David L. and {Patil}, Aarya A. and {Corrales}, Lia and {Brasseur}, C.~E. and {N{\"o}the}, Maximilian and {Donath}, Axel and {Tollerud}, Erik and {Morris}, Brett M. and {Ginsburg}, Adam and {Vaher}, Eero and {Weaver}, Benjamin A. and {Tocknell}, James and {Jamieson}, William and {van Kerkwijk}, Marten H. and {Robitaille}, Thomas P. and {Merry}, Bruce and {Bachetti}, Matteo and {G{\"u}nther}, H. Moritz and {Aldcroft}, Thomas L. and {Alvarado-Montes}, Jaime A. and {Archibald}, Anne M. and {B{\'o}di}, Attila and {Bapat}, Shreyas and {Barentsen}, Geert and {Baz{\'a}n}, Juanjo and {Biswas}, Manish and {Boquien}, M{\'e}d{\'e}ric and {Burke}, D.~J. and {Cara}, Daria and {Cara}, Mihai and {Conroy}, Kyle E. and {Conseil}, Simon and {Craig}, Matthew W. and {Cross}, Robert M. and {Cruz}, Kelle L. and {D'Eugenio}, Francesco and {Dencheva}, Nadia and {Devillepoix}, Hadrien A.~R. and {Dietrich}, J{\"o}rg P. and {Eigenbrot}, Arthur Davis and {Erben}, Thomas and {Ferreira}, Leonardo and {Foreman-Mackey}, Daniel and {Fox}, Ryan and {Freij}, Nabil and {Garg}, Suyog and {Geda}, Robel and {Glattly}, Lauren and {Gondhalekar}, Yash and {Gordon}, Karl D. and {Grant}, David and {Greenfield}, Perry and {Groener}, Austen M. and {Guest}, Steve and {Gurovich}, Sebastian and {Handberg}, Rasmus and {Hart}, Akeem and {Hatfield-Dodds}, Zac and {Homeier}, Derek and {Hosseinzadeh}, Griffin and {Jenness}, Tim and {Jones}, Craig K. and {Joseph}, Prajwel and {Kalmbach}, J. Bryce and {Karamehmetoglu}, Emir and {Ka{\l}uszy{\'n}ski}, Miko{\l}aj and {Kelley}, Michael S.~P. and {Kern}, Nicholas and {Kerzendorf}, Wolfgang E. and {Koch}, Eric W. and {Kulumani}, Shankar and {Lee}, Antony and {Ly}, Chun and {Ma}, Zhiyuan and {MacBride}, Conor and {Maljaars}, Jakob M. and {Muna}, Demitri and {Murphy}, N.~A. and {Norman}, Henrik and {O'Steen}, Richard and {Oman}, Kyle A. and {Pacifici}, Camilla and {Pascual}, Sergio and {Pascual-Granado}, J. and {Patil}, Rohit R. and {Perren}, Gabriel I. and {Pickering}, Timothy E. and {Rastogi}, Tanuj and {Roulston}, Benjamin R. and {Ryan}, Daniel F. and {Rykoff}, Eli S. and {Sabater}, Jose and {Sakurikar}, Parikshit and {Salgado}, Jes{\'u}s and {Sanghi}, Aniket and {Saunders}, Nicholas and {Savchenko}, Volodymyr and {Schwardt}, Ludwig and {Seifert-Eckert}, Michael and {Shih}, Albert Y. and {Jain}, Anany Shrey and {Shukla}, Gyanendra and {Sick}, Jonathan and {Simpson}, Chris and {Singanamalla}, Sudheesh and {Singer}, Leo P. and {Singhal}, Jaladh and {Sinha}, Manodeep and {Sip{\H{o}}cz}, Brigitta M. and {Spitler}, Lee R. and {Stansby}, David and {Streicher}, Ole and {{\v{S}}umak}, Jani and {Swinbank}, John D. and {Taranu}, Dan S. and {Tewary}, Nikita and {Tremblay}, Grant R. and {de Val-Borro}, Miguel and {Van Kooten}, Samuel J. and {Vasovi{\'c}}, Zlatan and {Verma}, Shresth and {de Miranda Cardoso}, Jos{\'e} Vin{\'\i}cius and {Williams}, Peter K.~G. and {Wilson}, Tom J. and {Winkel}, Benjamin and {Wood-Vasey}, W.~M. and {Xue}, Rui and {Yoachim}, Peter and {Zhang}, Chen and {Zonca}, Andrea and {Astropy Project Contributors}},
        title = "{The Astropy Project: Sustaining and Growing a Community-oriented Open-source Project and the Latest Major Release (v5.0) of the Core Package}",
      journal = {\apj},
         year = 2022,
        month = aug,
       volume = {935},
       number = {2},
          eid = {167},
        pages = {167},
          doi = {10.3847/1538-4357/ac7c74},
archivePrefix = {arXiv},
       eprint = {2206.14220},
 primaryClass = {astro-ph.IM},
       adsurl = {https://ui.adsabs.harvard.edu/abs/2022ApJ...935..167A}
}

@misc{2021zndo...4624996B,
       author = {{Bradley}, Larry and {Sip{\H{o}}cz}, Brigitta and {Robitaille}, Thomas and {Tollerud}, Erik and {Vin{\'\i}cius}, Z{\'e} and {Deil}, Christoph and {Barbary}, Kyle and {Wilson}, Tom J and {Busko}, Ivo and {Donath}, Axel and {G{\"u}nther}, Hans Moritz and {Cara}, Mihai and {Conseil}, Simon and {Bostroem}, Azalee and {Droettboom}, Michael and {Bray}, E.~M. and {Krachyon} and {Lim}, P.~L. and {Andersen Bratholm}, Lars and {Barentsen}, Geert and {Craig}, Matt and {Rathi}, Shivangee and {Pascual}, Sergio and {Perren}, Gabriel and {Georgiev}, Iskren Y. and {De Val-Borro}, Miguel and {Kerzendorf}, Wolfgang and {Bach}, Yoonsoo P. and {Quint}, Bruno and {Souchereau}, Harrison},
        title = "{astropy/photutils: 1.1.0}",
         year = 2021,
        month = mar,
          eid = {10.5281/zenodo.4624996},
          doi = {10.5281/zenodo.4624996},
      version = {1.1.0},
    publisher = {Zenodo},
       adsurl = {https://ui.adsabs.harvard.edu/abs/2021zndo...4624996B}
}

@ARTICLE{2020ApJ...889L..30L,
       author = {{Lin}, Hsing Wen and {Lee}, Chien-Hsiu and {Gerdes}, D.~W. and {Adams}, Fred C. and {Becker}, Juliette and {Napier}, Kevin and {Markwardt}, Larissa},
        title = "{Detection of Diatomic Carbon in 2I/Borisov}",
      journal = {\apjl},
         year = 2020,
        month = feb,
       volume = {889},
       number = {2},
          eid = {L30},
        pages = {L30},
          doi = {10.3847/2041-8213/ab6bd9},
archivePrefix = {arXiv},
       eprint = {1912.06161},
 primaryClass = {astro-ph.EP},
       adsurl = {https://ui.adsabs.harvard.edu/abs/2020ApJ...889L..30L}
}

@ARTICLE{2020NatMe..17..261V,
       author = {{Virtanen}, Pauli and {Gommers}, Ralf and {Oliphant}, Travis E. and {Haberland}, Matt and {Reddy}, Tyler and {Cournapeau}, David and {Burovski}, Evgeni and {Peterson}, Pearu and {Weckesser}, Warren and {Bright}, Jonathan and {van der Walt}, St{\'e}fan J. and {Brett}, Matthew and {Wilson}, Joshua and {Millman}, K. Jarrod and {Mayorov}, Nikolay and {Nelson}, Andrew R.~J. and {Jones}, Eric and {Kern}, Robert and {Larson}, Eric and {Carey}, C.~J. and {Polat}, {\.I}lhan and {Feng}, Yu and {Moore}, Eric W. and {VanderPlas}, Jake and {Laxalde}, Denis and {Perktold}, Josef and {Cimrman}, Robert and {Henriksen}, Ian and {Quintero}, E.~A. and {Harris}, Charles R. and {Archibald}, Anne M. and {Ribeiro}, Ant{\^o}nio H. and {Pedregosa}, Fabian and {van Mulbregt}, Paul and {SciPy 1.  0 Contributors}},
        title = "{SciPy 1.0: fundamental algorithms for scientific computing in Python}",
      journal = {Nature Medicine},
         year = 2020,
        month = feb,
       volume = {17},
        pages = {261-272},
          doi = {10.1038/s41592-019-0686-2},
archivePrefix = {arXiv},
       eprint = {1907.10121},
 primaryClass = {cs.MS},
       adsurl = {https://ui.adsabs.harvard.edu/abs/2020NatMe..17..261V}
}

@ARTICLE{2020NatAs...4...53G,
       author = {{Guzik}, Piotr and {Drahus}, Micha{\l} and {Rusek}, Krzysztof and {Waniak}, Wac{\l}aw and {Cannizzaro}, Giacomo and {Pastor-Marazuela}, In{\'e}s},
        title = "{Initial characterization of interstellar comet 2I/Borisov}",
      journal = {Nature Astronomy},
         year = 2020,
        month = jan,
       volume = {4},
        pages = {53-57},
          doi = {10.1038/s41550-019-0931-8},
archivePrefix = {arXiv},
       eprint = {1909.05851},
 primaryClass = {astro-ph.EP},
       adsurl = {https://ui.adsabs.harvard.edu/abs/2020NatAs...4...53G}
}

@ARTICLE{2019ApJ...886L..29J,
       author = {{Jewitt}, David and {Luu}, Jane},
        title = "{Initial Characterization of Interstellar Comet 2I/2019 Q4 (Borisov)}",
      journal = {\apjl},
         year = 2019,
        month = dec,
       volume = {886},
       number = {2},
          eid = {L29},
        pages = {L29},
          doi = {10.3847/2041-8213/ab530b},
archivePrefix = {arXiv},
       eprint = {1910.02547},
 primaryClass = {astro-ph.EP},
       adsurl = {https://ui.adsabs.harvard.edu/abs/2019ApJ...886L..29J}
}

@ARTICLE{2019ApJ...885L...9F,
       author = {{Fitzsimmons}, Alan and {Hainaut}, Olivier and {Meech}, Karen J. and {Jehin}, Emmanuel and {Moulane}, Youssef and {Opitom}, Cyrielle and {Yang}, Bin and {Keane}, Jacqueline V. and {Kleyna}, Jan T. and {Micheli}, Marco and {Snodgrass}, Colin},
        title = "{Detection of CN Gas in Interstellar Object 2I/Borisov}",
      journal = {\apjl},
         year = 2019,
        month = nov,
       volume = {885},
       number = {1},
          eid = {L9},
        pages = {L9},
          doi = {10.3847/2041-8213/ab49fc},
archivePrefix = {arXiv},
       eprint = {1909.12144},
 primaryClass = {astro-ph.EP},
       adsurl = {https://ui.adsabs.harvard.edu/abs/2019ApJ...885L...9F}
}

@ARTICLE{2018AJ....156..123A,
       author = {{Astropy Collaboration} and {Price-Whelan}, A.~M. and {Sip{\H{o}}cz}, B.~M. and {G{\"u}nther}, H.~M. and {Lim}, P.~L. and {Crawford}, S.~M. and {Conseil}, S. and {Shupe}, D.~L. and {Craig}, M.~W. and {Dencheva}, N. and {Ginsburg}, A. and {VanderPlas}, J.~T. and {Bradley}, L.~D. and {P{\'e}rez-Su{\'a}rez}, D. and {de Val-Borro}, M. and {Aldcroft}, T.~L. and {Cruz}, K.~L. and {Robitaille}, T.~P. and {Tollerud}, E.~J. and {Ardelean}, C. and {Babej}, T. and {Bach}, Y.~P. and {Bachetti}, M. and {Bakanov}, A.~V. and {Bamford}, S.~P. and {Barentsen}, G. and {Barmby}, P. and {Baumbach}, A. and {Berry}, K.~L. and {Biscani}, F. and {Boquien}, M. and {Bostroem}, K.~A. and {Bouma}, L.~G. and {Brammer}, G.~B. and {Bray}, E.~M. and {Breytenbach}, H. and {Buddelmeijer}, H. and {Burke}, D.~J. and {Calderone}, G. and {Cano Rodr{\'\i}guez}, J.~L. and {Cara}, M. and {Cardoso}, J.~V.~M. and {Cheedella}, S. and {Copin}, Y. and {Corrales}, L. and {Crichton}, D. and {D'Avella}, D. and {Deil}, C. and {Depagne}, {\'E}. and {Dietrich}, J.~P. and {Donath}, A. and {Droettboom}, M. and {Earl}, N. and {Erben}, T. and {Fabbro}, S. and {Ferreira}, L.~A. and {Finethy}, T. and {Fox}, R.~T. and {Garrison}, L.~H. and {Gibbons}, S.~L.~J. and {Goldstein}, D.~A. and {Gommers}, R. and {Greco}, J.~P. and {Greenfield}, P. and {Groener}, A.~M. and {Grollier}, F. and {Hagen}, A. and {Hirst}, P. and {Homeier}, D. and {Horton}, A.~J. and {Hosseinzadeh}, G. and {Hu}, L. and {Hunkeler}, J.~S. and {Ivezi{\'c}}, {\v{Z}}. and {Jain}, A. and {Jenness}, T. and {Kanarek}, G. and {Kendrew}, S. and {Kern}, N.~S. and {Kerzendorf}, W.~E. and {Khvalko}, A. and {King}, J. and {Kirkby}, D. and {Kulkarni}, A.~M. and {Kumar}, A. and {Lee}, A. and {Lenz}, D. and {Littlefair}, S.~P. and {Ma}, Z. and {Macleod}, D.~M. and {Mastropietro}, M. and {McCully}, C. and {Montagnac}, S. and {Morris}, B.~M. and {Mueller}, M. and {Mumford}, S.~J. and {Muna}, D. and {Murphy}, N.~A. and {Nelson}, S. and {Nguyen}, G.~H. and {Ninan}, J.~P. and {N{\"o}the}, M. and {Ogaz}, S. and {Oh}, S. and {Parejko}, J.~K. and {Parley}, N. and {Pascual}, S. and {Patil}, R. and {Patil}, A.~A. and {Plunkett}, A.~L. and {Prochaska}, J.~X. and {Rastogi}, T. and {Reddy Janga}, V. and {Sabater}, J. and {Sakurikar}, P. and {Seifert}, M. and {Sherbert}, L.~E. and {Sherwood-Taylor}, H. and {Shih}, A.~Y. and {Sick}, J. and {Silbiger}, M.~T. and {Singanamalla}, S. and {Singer}, L.~P. and {Sladen}, P.~H. and {Sooley}, K.~A. and {Sornarajah}, S. and {Streicher}, O. and {Teuben}, P. and {Thomas}, S.~W. and {Tremblay}, G.~R. and {Turner}, J.~E.~H. and {Terr{\'o}n}, V. and {van Kerkwijk}, M.~H. and {de la Vega}, A. and {Watkins}, L.~L. and {Weaver}, B.~A. and {Whitmore}, J.~B. and {Woillez}, J. and {Zabalza}, V. and {Astropy Contributors}},
        title = "{The Astropy Project: Building an Open-science Project and Status of the v2.0 Core Package}",
      journal = {\aj},
         year = 2018,
        month = sep,
       volume = {156},
       number = {3},
          eid = {123},
        pages = {123},
          doi = {10.3847/1538-3881/aabc4f},
archivePrefix = {arXiv},
       eprint = {1801.02634},
 primaryClass = {astro-ph.IM},
       adsurl = {https://ui.adsabs.harvard.edu/abs/2018AJ....156..123A}
}

@ARTICLE{2018Natur.559..223M,
       author = {{Micheli}, Marco and {Farnocchia}, Davide and {Meech}, Karen J. and {Buie}, Marc W. and {Hainaut}, Olivier R. and {Prialnik}, Dina and {Sch{\"o}rghofer}, Norbert and {Weaver}, Harold A. and {Chodas}, Paul W. and {Kleyna}, Jan T. and {Weryk}, Robert and {Wainscoat}, Richard J. and {Ebeling}, Harald and {Keane}, Jacqueline V. and {Chambers}, Kenneth C. and {Koschny}, Detlef and {Petropoulos}, Anastassios E.},
        title = "{Non-gravitational acceleration in the trajectory of 1I/2017 U1 ('Oumuamua)}",
      journal = {\nat},
         year = 2018,
        month = jun,
       volume = {559},
        pages = {223-226},
          doi = {10.1038/s41586-018-0254-4},
       adsurl = {https://ui.adsabs.harvard.edu/abs/2018Natur.559..223M}
}

@ARTICLE{2018ApJ...852L...2B,
       author = {{Bolin}, Bryce T. and {Weaver}, Harold A. and {Fernandez}, Yanga R. and {Lisse}, Carey M. and {Huppenkothen}, Daniela and {Jones}, R. Lynne and {Juri{\'c}}, Mario and {Moeyens}, Joachim and {Schambeau}, Charles A. and {Slater}, Colin. T. and {Ivezi{\'c}}, {\v{Z}}eljko and {Connolly}, Andrew J.},
        title = "{APO Time-resolved Color Photometry of Highly Elongated Interstellar Object 1I/{\textquoteleft}Oumuamua}",
      journal = {\apjl},
         year = 2018,
        month = jan,
       volume = {852},
       number = {1},
          eid = {L2},
        pages = {L2},
          doi = {10.3847/2041-8213/aaa0c9},
archivePrefix = {arXiv},
       eprint = {1711.04927},
 primaryClass = {astro-ph.EP},
       adsurl = {https://ui.adsabs.harvard.edu/abs/2018ApJ...852L...2B}
}

@ARTICLE{2017ApJ...851L..38B,
       author = {{Bannister}, Michele T. and {Schwamb}, Megan E. and {Fraser}, Wesley C. and {Marsset}, Michael and {Fitzsimmons}, Alan and {Benecchi}, Susan D. and {Lacerda}, Pedro and {Pike}, Rosemary E. and {Kavelaars}, J.~J. and {Smith}, Adam B. and {Stewart}, Sunny O. and {Wang}, Shiang-Yu and {Lehner}, Matthew J.},
        title = "{Col-OSSOS: Colors of the Interstellar Planetesimal 1I/{\textquoteleft}Oumuamua}",
      journal = {\apjl},
         year = 2017,
        month = dec,
       volume = {851},
       number = {2},
          eid = {L38},
        pages = {L38},
          doi = {10.3847/2041-8213/aaa07c},
archivePrefix = {arXiv},
       eprint = {1711.06214},
 primaryClass = {astro-ph.EP},
       adsurl = {https://ui.adsabs.harvard.edu/abs/2017ApJ...851L..38B}
}

@ARTICLE{2017Natur.552..378M,
       author = {{Meech}, Karen J. and {Weryk}, Robert and {Micheli}, Marco and {Kleyna}, Jan T. and {Hainaut}, Olivier R. and {Jedicke}, Robert and {Wainscoat}, Richard J. and {Chambers}, Kenneth C. and {Keane}, Jacqueline V. and {Petric}, Andreea and {Denneau}, Larry and {Magnier}, Eugene and {Berger}, Travis and {Huber}, Mark E. and {Flewelling}, Heather and {Waters}, Chris and {Schunova-Lilly}, Eva and {Chastel}, Serge},
        title = "{A brief visit from a red and extremely elongated interstellar asteroid}",
      journal = {\nat},
         year = 2017,
        month = dec,
       volume = {552},
       number = {7685},
        pages = {378-381},
          doi = {10.1038/nature25020},
       adsurl = {https://ui.adsabs.harvard.edu/abs/2017Natur.552..378M}
}

@ARTICLE{2017MNRAS.469S.475R,
       author = {{Rosenbush}, Vera K. and {Ivanova}, Oleksandra V. and {Kiselev}, Nikolai N. and {Kolokolova}, Ludmilla O. and {Afanasiev}, Viktor L.},
        title = "{Spatial variations of brightness, colour and polarization of dust in comet 67P/Churyumov-Gerasimenko}",
      journal = {\mnras},
         year = 2017,
        month = jul,
       volume = {469},
        pages = {S475-S491},
          doi = {10.1093/mnras/stx2003},
       adsurl = {https://ui.adsabs.harvard.edu/abs/2017MNRAS.469S.475R}
}

@ARTICLE{2015AJ....150..201J,
       author = {{Jewitt}, David},
        title = "{Color Systematics of Comets and Related Bodies}",
      journal = {\aj},
         year = 2015,
        month = dec,
       volume = {150},
       number = {6},
          eid = {201},
        pages = {201},
          doi = {10.1088/0004-6256/150/6/201},
archivePrefix = {arXiv},
       eprint = {1510.07069},
 primaryClass = {astro-ph.EP},
       adsurl = {https://ui.adsabs.harvard.edu/abs/2015AJ....150..201J}
}

@ARTICLE{2013A&A...558A..33A,
       author = {{Astropy Collaboration} and {Robitaille}, Thomas P. and {Tollerud}, Erik J. and {Greenfield}, Perry and {Droettboom}, Michael and {Bray}, Erik and {Aldcroft}, Tom and {Davis}, Matt and {Ginsburg}, Adam and {Price-Whelan}, Adrian M. and {Kerzendorf}, Wolfgang E. and {Conley}, Alexander and {Crighton}, Neil and {Barbary}, Kyle and {Muna}, Demitri and {Ferguson}, Henry and {Grollier}, Fr{\'e}d{\'e}ric and {Parikh}, Madhura M. and {Nair}, Prasanth H. and {Unther}, Hans M. and {Deil}, Christoph and {Woillez}, Julien and {Conseil}, Simon and {Kramer}, Roban and {Turner}, James E.~H. and {Singer}, Leo and {Fox}, Ryan and {Weaver}, Benjamin A. and {Zabalza}, Victor and {Edwards}, Zachary I. and {Azalee Bostroem}, K. and {Burke}, D.~J. and {Casey}, Andrew R. and {Crawford}, Steven M. and {Dencheva}, Nadia and {Ely}, Justin and {Jenness}, Tim and {Labrie}, Kathleen and {Lim}, Pey Lian and {Pierfederici}, Francesco and {Pontzen}, Andrew and {Ptak}, Andy and {Refsdal}, Brian and {Servillat}, Mathieu and {Streicher}, Ole},
        title = "{Astropy: A community Python package for astronomy}",
      journal = {\aap},
         year = 2013,
        month = oct,
       volume = {558},
          eid = {A33},
        pages = {A33},
          doi = {10.1051/0004-6361/201322068},
archivePrefix = {arXiv},
       eprint = {1307.6212},
 primaryClass = {astro-ph.IM},
       adsurl = {https://ui.adsabs.harvard.edu/abs/2013A&A...558A..33A}
}

@ARTICLE{2013MNRAS.430..330H,
       author = {{H{\"a}u{\ss}ler}, Boris and {Bamford}, Steven P. and {Vika}, Marina and {Rojas}, Alex L. and {Barden}, Marco and {Kelvin}, Lee S. and {Alpaslan}, Mehmet and {Robotham}, Aaron S.~G. and {Driver}, Simon P. and {Baldry}, I.~K. and {Brough}, Sarah and {Hopkins}, Andrew M. and {Liske}, Jochen and {Nichol}, Robert C. and {Popescu}, Cristina C. and {Tuffs}, Richard J.},
        title = "{MegaMorph - multiwavelength measurement of galaxy structure: complete S{\'e}rsic profile information from modern surveys}",
      journal = {\mnras},
         year = 2013,
        month = mar,
       volume = {430},
       number = {1},
        pages = {330-369},
          doi = {10.1093/mnras/sts633},
archivePrefix = {arXiv},
       eprint = {1212.3332},
 primaryClass = {astro-ph.CO},
       adsurl = {https://ui.adsabs.harvard.edu/abs/2013MNRAS.430..330H}
}

@misc{2013ascl.soft01001B,
       author = {{Bertin}, Emmanuel},
        title = "{PSFEx: Point Spread Function Extractor}",
 howpublished = {Astrophysics Source Code Library, record ascl:1301.001},
         year = 2013,
        month = jan,
          eid = {ascl:1301.001},
archivePrefix = {ascl},
       eprint = {1301.001},
       adsurl = {https://ui.adsabs.harvard.edu/abs/2013ascl.soft01001B}
}

@misc{2010ascl.soft10068B,
       author = {{Bertin}, Emmanuel},
        title = "{SWarp: Resampling and Co-adding FITS Images Together}",
 howpublished = {Astrophysics Source Code Library, record ascl:1010.068},
         year = 2010,
        month = oct,
          eid = {ascl:1010.068},
archivePrefix = {ascl},
       eprint = {1010.068},
       adsurl = {https://ui.adsabs.harvard.edu/abs/2010ascl.soft10068B}
}

@ARTICLE{2010AJ....139.2097P,
       author = {{Peng}, Chien Y. and {Ho}, Luis C. and {Impey}, Chris D. and {Rix}, Hans-Walter},
        title = "{Detailed Decomposition of Galaxy Images. II. Beyond Axisymmetric Models}",
      journal = {\aj},
         year = 2010,
        month = jun,
       volume = {139},
       number = {6},
        pages = {2097-2129},
          doi = {10.1088/0004-6256/139/6/2097},
archivePrefix = {arXiv},
       eprint = {0912.0731},
 primaryClass = {astro-ph.CO},
       adsurl = {https://ui.adsabs.harvard.edu/abs/2010AJ....139.2097P}
}

@ARTICLE{2009Icar..202..160D,
       author = {{DeMeo}, Francesca E. and {Binzel}, Richard P. and {Slivan}, Stephen M. and {Bus}, Schelte J.},
        title = "{An extension of the Bus asteroid taxonomy into the near-infrared}",
      journal = {\icarus},
         year = 2009,
        month = jul,
       volume = {202},
       number = {1},
        pages = {160-180},
          doi = {10.1016/j.icarus.2009.02.005},
       adsurl = {https://ui.adsabs.harvard.edu/abs/2009Icar..202..160D}
}

@INPROCEEDINGS{2006ASPC..351..112B,
       author = {{Bertin}, E.},
        title = "{Automatic Astrometric and Photometric Calibration with SCAMP}",
    booktitle = {Astronomical Data Analysis Software and Systems XV},
         year = 2006,
       editor = {{Gabriel}, C. and {Arviset}, C. and {Ponz}, D. and {Enrique}, S.},
       series = {Astronomical Society of the Pacific Conference Series},
       volume = {351},
        month = jul,
        pages = {112},
       adsurl = {https://ui.adsabs.harvard.edu/abs/2006ASPC..351..112B}
}

@INCOLLECTION{2004come.book..425F,
       author = {{Feldman}, P.~D. and {Cochran}, A.~L. and {Combi}, M.~R.},
        title = "{Spectroscopic investigations of fragment species in the coma}",
    booktitle = {Comets II},
         year = 2004,
       editor = {{Festou}, Michel C. and {Keller}, H. Uwe and {Weaver}, Harold A.},
        pages = {425},
       adsurl = {https://ui.adsabs.harvard.edu/abs/2004come.book..425F}
}

@ARTICLE{2002AJ....124..266P,
       author = {{Peng}, Chien Y. and {Ho}, Luis C. and {Impey}, Chris D. and {Rix}, Hans-Walter},
        title = "{Detailed Structural Decomposition of Galaxy Images}",
      journal = {\aj},
         year = 2002,
        month = jul,
       volume = {124},
       number = {1},
        pages = {266-293},
          doi = {10.1086/340952},
archivePrefix = {arXiv},
       eprint = {astro-ph/0204182},
 primaryClass = {astro-ph},
       adsurl = {https://ui.adsabs.harvard.edu/abs/2002AJ....124..266P}
}

@ARTICLE{2002AJ....123.1039J,
       author = {{Jewitt}, David C.},
        title = "{From Kuiper Belt Object to Cometary Nucleus: The Missing Ultrared Matter}",
      journal = {\aj},
         year = 2002,
        month = feb,
       volume = {123},
       number = {2},
        pages = {1039-1049},
          doi = {10.1086/338692},
       adsurl = {https://ui.adsabs.harvard.edu/abs/2002AJ....123.1039J}
}

@INPROCEEDINGS{1996DPS....28.2504G,
       author = {{Giorgini}, J.~D. and {Yeomans}, D.~K. and {Chamberlin}, A.~B. and {Chodas}, P.~W. and {Jacobson}, R.~A. and {Keesey}, M.~S. and {Lieske}, J.~H. and {Ostro}, S.~J. and {Standish}, E.~M. and {Wimberly}, R.~N.},
        title = "{JPL's On-Line Solar System Data Service}",
    booktitle = {AAS/Division for Planetary Sciences Meeting Abstracts \#28},
         year = 1996,
       series = {AAS/Division for Planetary Sciences Meeting Abstracts},
       volume = {28},
        month = sep,
          eid = {25.04},
        pages = {25.04},
       adsurl = {https://ui.adsabs.harvard.edu/abs/1996DPS....28.2504G}
}

@ARTICLE{1996A&AS..117..393B,
       author = {{Bertin}, E. and {Arnouts}, S.},
        title = "{SExtractor: Software for source extraction.}",
      journal = {\aaps},
         year = 1996,
        month = jun,
       volume = {117},
        pages = {393-404},
          doi = {10.1051/aas:1996164},
       adsurl = {https://ui.adsabs.harvard.edu/abs/1996A&AS..117..393B}
}

@ARTICLE{1995Icar..118..223A,
       author = {{A'Hearn}, Michael F. and {Millis}, Robert C. and {Schleicher}, David O. and {Osip}, David J. and {Birch}, Peter V.},
        title = "{The ensemble properties of comets: Results from narrowband photometry of 85 comets, 1976-1992.}",
      journal = {\icarus},
         year = 1995,
        month = dec,
       volume = {118},
       number = {2},
        pages = {223-270},
          doi = {10.1006/icar.1995.1190},
       adsurl = {https://ui.adsabs.harvard.edu/abs/1995Icar..118..223A}
}

@INPROCEEDINGS{1989aste.conf..524B,
       author = {{Bowell}, Edward and {Hapke}, Bruce and {Domingue}, Deborah and {Lumme}, Kari and {Peltoniemi}, Jouni and {Harris}, Alan W.},
        title = "{Application of photometric models to asteroids.}",
    booktitle = {Asteroids II},
         year = 1989,
       editor = {{Binzel}, Richard P. and {Gehrels}, Tom and {Matthews}, Mildred Shapley},
        month = jan,
        pages = {524-556},
       adsurl = {https://ui.adsabs.harvard.edu/abs/1989aste.conf..524B}
}

@ARTICLE{1984AJ.....89..579A,
       author = {{A'Hearn}, M.~F. and {Schleicher}, D.~G. and {Millis}, R.~L. and {Feldman}, P.~D. and {Thompson}, D.~T.},
        title = "{Comet Bowell 1980b}",
      journal = {\aj},
         year = 1984,
        month = apr,
       volume = {89},
        pages = {579-591},
          doi = {10.1086/113552},
       adsurl = {https://ui.adsabs.harvard.edu/abs/1984AJ.....89..579A}
}

@ARTICLE{1957BSRSL..43..740H,
       author = {{Haser}, L.},
        title = "{Distribution d'intensit{\'e} dans la t{\^e}te d'une com{\`e}te}",
      journal = {Bulletin de la Societe Royale des Sciences de Liege},
         year = 1957,
        month = jan,
       volume = {43},
        pages = {740-750},
       adsurl = {https://ui.adsabs.harvard.edu/abs/1957BSRSL..43..740H}
}

@ARTICLE{2025arXiv251207318L,
       author = {{Lisse}, Carey M. and {Bach}, Yoonsoo P. and {Crill}, Brendan P. and {Korngut}, Phil M. and {Cukierman}, Ari J. and {Bryan}, Sean A. and {Cooray}, Asantha and {Dowell}, C. Darren and {Werner}, Michael W. and {Hora}, Joseph L. and {Rustamkulov}, Zafar and {Lee}, Jeong-Eun and {Lim}, Bumhoo and {Fernandez}, Y.~R. and {Tolls}, Volker and {Reach}, W.~T. and {Dor{\'e}}, O. and {Zemcov}, Michael and {Bock}, James J. and {Cheng}, Yun-Ting and {Champagne}, C. and {Choi}, Seungwon and {Connelley}, M. and {Emery}, J.~P. and {Everett}, Spencer and {Faisst}, Andreas L. and {Geem}, Jooyeon and {Hui}, Howard and {Ishiguro}, Masateru and {Jin}, Sunho and {Jo}, Hangbin and {Mahlke}, Max and {Masters}, Daniel C. and {Melnick}, Gary J. and {Nguyen}, Chi H. and {Paladini}, Roberta and {Sitko}, M.~L. and {Yang}, Yujin},
        title = "{SPHEREx Pre-Perihelion Mapping of $\mathrm{H_2O}$, $\mathrm{CO_2}$, and $\mathrm{CO}$ in Interstellar Object 3I/ATLAS}",
      journal = {arXiv e-prints},
         year = 2025,
        month = dec,
          eid = {arXiv:2512.07318},
        pages = {arXiv:2512.07318},
          doi = {10.48550/arXiv.2512.07318},
archivePrefix = {arXiv},
       eprint = {2512.07318},
 primaryClass = {astro-ph.EP},
       adsurl = {https://ui.adsabs.harvard.edu/abs/2025arXiv251207318L}
}

@ARTICLE{2026MNRAS.tmp...52C,
       author = {{Coulson}, Iain M. and {Kuan}, Yi-Jehng and {Charnley}, Steven B. and {Cordiner}, Martin A. and {Chuang}, Yo-Ling and {Lee}, Yueh-Ning and {Lin}, Min-Kai and {Milam}, Stefanie N. and {Pimpanuwat}, Bannawit and {Roth}, Nathan X. and {{\.Z}{\'o}{\l}towski}, Micha{\l}},
        title = "{JCMT detection of HCN emission from 3I/ATLAS at 2.1 AU}",
      journal = {\mnras},
         year = 2026,
        month = jan,
          doi = {10.1093/mnras/stag063},
archivePrefix = {arXiv},
       eprint = {2510.02817},
 primaryClass = {astro-ph.EP},
       adsurl = {https://ui.adsabs.harvard.edu/abs/2026MNRAS.tmp...52C}
}

@misc{paek2025_gppy_gpu,
  author       = {Paek, Gregory S. H.},
  title        = {gppy-gpu v1.0.0: A GPU-Accelerated Imaging Pipeline for the 7DT/7DS},
  year         = {2025},
  publisher    = {Zenodo},
  version      = {v1.0.0},
  doi          = {10.5281/zenodo. 17065902},
  url          = {https://doi.org/10.5281/zenodo. 17065902},
  note         = {Computer software}
}
\bibliographystyle{aasjournalv7}



\end{document}